\documentclass[
 reprint,
 amsmath,amssymb,
 aps, pra, floatfix
]{revtex4-2}
\usepackage[colorlinks]{hyperref}
\hypersetup{
    colorlinks,
    linkcolor={blue},
    citecolor={blue},
    urlcolor={blue}
}
\usepackage{graphicx}
\usepackage{dcolumn}
\usepackage{bm}
\usepackage{xcolor}

\begin{document}
\preprint{APS/123-QED}

\title{Gaussian tripartite entanglement in the simultaneous measurement of\\  position and momentum}

\author{J. A. Mendoza-Fierro}
\email{julio.mendoza@alumno.buap.mx}
\author{L. M. Ar\' evalo Aguilar}
\email{larevalo@fcfm.buap.mx}
\affiliation{Facultad de Ciencias Físico Matemáticas, Benemérita Universidad Autónoma de Puebla.}
\date{\today}

\begin{abstract}
In this work, we prove the generation of genuine tripartite continuous-variable entanglement in the unitary dynamics of the simultaneous measurement process of position and momentum observables raised by Arthurs and Kelly, considering a measurement configuration where the system under examination is a rotated, displaced, and squeezed vacuum state. Under these assumptions, the measurement configuration is entirely described by a Gaussian state. Then, through the positive partial transpose criterion (PPT), we certify genuine tripartite entanglement by testing the non-separability of the three $\left(1~ \text{vs}~2\right)$-mode bipartitions of the system.  This process allows us to classify the qualitative properties of the entanglement in the category of fully inseparable Gaussian states according to the classification exposed in [Giedke et al., \href{https://link.aps.org/doi/10.1103/PhysRevA.64.052303}{Phys. Rev. A \textbf{64}, 052303 (2001)}]. Besides, we determine the quantitative entanglement properties of the system using the residual tripartite R{\'e}nyi-2 entanglement as a quantifier measure.   
\end{abstract}
\maketitle
\section{Introduction}
Quantum entanglement is the most distinctive trait of quantum mechanical compound systems. The first mathematically rigorous manifestation of this concept comes back from the von Neumann proposal for the standard model of a quantum measurement \cite{Neumann1955}; however, were Einstein, Podolsky, and Rosen who put in context one of the implications of the quantum entanglement through the existence of non-locality in a pure non-factorizable two-particle system \cite{Einstein1935}. The manifestation of a global state for a multipartite system which posses non-classical correlations between the subsystems was called by Schrödinger as ``Verschränkung'' (entanglement) \cite{Schrodinger1935}. Therefore, the quantum entanglement evolved from studying the completeness of quantum theory up to be posed as the fundamental ingredient for practical tasks which would be impossible in the classical domain; namely, quantum cryptography \cite{Bennett1984, Ekert1991quantum}, quantum dense coding \cite{Bennett1992}, quantum teleportation \cite{Bennett1993}, and quantum computation \cite{Feynman1982, *Deutsch1985, *Shor1995}, establishing the basis for the field of quantum information \cite{Lo1998, Nielsen2000, Bergou2013}.

Nowadays, there exists considerable interest in the entanglement of multipartite systems, that is, those which consider more than two subsystems.  Its usefulness lies in the potential benefits for the quantum information processing; for example, the creation of entanglement between many atoms or ions \cite{Leibfried2005, *Haffner2005, *Monz2011} is useful for fault-tolerant quantum computing \cite{Shor1996, Steane2005, Knill2005}, high-precision measurements using matter-wave interference \cite{Gross2010} and the fundamental investigation of the quantum to classical regime transition \cite{Myatt2000, *Deleglise2008, *Barreiro2010}. In the case of many entangled photons \cite{Pan2000, *Bourennane2004, *Lu2007, *Wieczorek2009, *Papp2009, *Deng2011, *Yao2012}, it allows entanglement verification in quantum networks \cite{Mccutcheon2016}, generation of cluster states \cite{Schwartz2016} and error correction protocols \cite{Deng2011b, *Sheng2009, *Gao2017}. Further, we can generate multipartite entanglement in the continuous variable (CV) regime by using modes of the electromagnetic field \cite{Aoki2003, Coelho2009, Armstrong2012}, which find applicability in teleportation networks \cite{Van2000} and multiuser quantum channel for telecloning \cite{Van2001a}. In the present paper, we will be concerned only with the entanglement generation in the CV context.

In particular, the multipartite entanglement in CV systems constitutes a valuable resource for fundamental studies and practical applications; the archetypical paradigm is represented by the Gaussian states, which are those who present a Gaussian profile in the quantum phase space. The Gaussian states are referent from the optical domain, like the coherent, squeezed, thermal, and vacuum states, however, also living in other physical scenarios as in trapped ions \cite{Serafini2009, Serafini2009b, Alonso2013},  nanomechanical resonators \cite{Rabl2004, Wollman2015} and optical cavities \cite{Slusher1985, Reid1986}. The useful properties of Gaussian states include easy experimental generation, simple mathematical description, and accessible resources for creating quantum entanglement; therefore, they are the preferred systems for CV quantum informational tasks. Besides, the Gaussian states are the test-bed for fundamental studies in CV systems as non-separability \cite{Giedke2001}, entanglement sharing \cite{Adesso2006, Hiroshima2007}, and entanglement and correlation measures \cite{Adesso2006, Adesso2006a, Adesso2014}. 

The first case of multipartite Gaussian entangled system is given by the three-partie scenario. A particular example is the CV GHZ states \cite{Van2000, Van2001}; these states can be generated by combining two orthogonally squeezed vacuum states through a beam splitter and then interfering one of the outputs arms with a third squeezed state in a second beam splitter \cite{Van2003} (see Fig. 2 of \cite{Teh2014}). The CV GHZ states have been generated experimentally in Ref. \cite{Aoki2003}. Besides, it is possible to create tripartite polarization-entangled states of bright optical beams by taking advantage of the generating configuration of the CV GHZ states plus interfering coherent beams on a beam splitter for each of the three entangled output modes (see Fig. 1 of \cite{Wu2016}); the resulting states are associated with spin states of atomic ensembles. An alternative scenario is the CV EPR-type state, where the third squeezed mode of the CV GHZ configuration is replaced by a coherent vacuum state (see Fig. 3 of \cite{Teh2014}); these states were generated in Refs. \cite{Armstrong2012, Armstrong2015}. Then, the establishment of genuine tripartite entanglement in Gaussian states gives a wide variety of applications in quantum communication processes, offering the direct possibility to improve the new wave of quantum technologies which will make up the second quantum revolution \cite{Deutsch2020}. 

In this work, we establish the simultaneous measurement process of the position and momentum observables of an arbitrary system raised by Arthurs and Kelly \cite{Arthurs1965} as a generator of genuine tripartite Gaussian entanglement when the system under measurement becomes a particular Gaussian system. We prove the non-separability of each of the three mode bipartitions of the system; hence, since the original proposal considers only pure states, we certify the complete entanglement among the three modes of the system.  Besides, given the exhaustive classification for three-mode Gaussian systems in function of the separability of their bipartitions \cite{Giedke2001}, we conclude that the Gaussian state describing the measurement process is fully inseparable. Moreover, we quantify the entanglement of the system using the residual tripartite R{\'e}nyi-2 entanglement \cite{Adesso2014}.

Throughout this article, unless otherwise stated, we will use units of $\hbar=2$. Besides, the structure is as follows: In Sec. \ref{Sec. 2} we review the measurement process of Arthurs and Kelly; also, we establish the measurement setup and the dynamics of the system. In Sec. \ref{Sec. 3} we review some fundamental definitions for Gaussian states. Moreover, we test the non-separability of the three mode-bipartitions of the measurement configuration; besides, we quantify the tripartite entanglement contained in the system. We close with the conclusions in \ref{Sec. 4}.

\section{The simultaneous measurement of position and momentum} \label{Sec. 2}
As generalization of the standard von Neumann measurement model \cite{Neumann1955}, Arthurs and Kelly \cite{Arthurs1965} proposed a measurement scheme in which two simultaneously diagonalizable (commuting) quantum observables associated with a pair of detectors of a measuring device are coupled with the position and momentum observables of an arbitrary system. This process is achieved through the following interaction Hamiltonian
\begin{equation} 
\hat{H}_{\text{int}}= \kappa_{1} \hat{x}
_{3} \hat{p}_{1} + \kappa_{2} \hat{p}_{3} \hat{p}_{2}; \label{eq:1}
\end{equation} 												
in the subsequent, the labels 1 and 2 will refer to the variables of the first and second detector, respectively, while label 3 will represent the the system under measurement. The Hamiltonian, Eq. \eqref{eq:1}, implies that the positions of the pointers will be displaced, within a proportional factor of $\kappa_{j} ~(j=1,2)$, by the canonical pair of the system under observation. Then, by repeating the measurement process on a large number of identically prepared systems, one can obtain information about the initial conjugate pair on interest within some degree of error which results from the inherent noise carried by the measurement process \cite{Arthurs1965}. 

In particular, this mechanism to transfer information is possible due to the entanglement arising before the readings of the detectors, which is an exclusively quantum feature in the interaction stage of any measurement process, where the quantum state of the system under observation becomes correlated with the state of the measuring device \cite{Peres1986, Patekar2019}. Then, under the so-called joint unbiasedness condition, Arthurs and Kelly show that the marginal statistics of the measurement outputs obey the uncertainty relation (units of $\hbar=1$) 
\begin{equation}
\delta_{\hat{x}_{3}}\delta_{\hat{p}_{3}} \geq 1,\label{eq:2}
\end{equation}
what is essentially different from the well-known Heisenberg uncertainty relation, which, rather, quantifies the noise of separate ideal single measurements on separated members of an ensemble of identically prepared systems \cite{She1966}. Instead, in Eq. \eqref{eq:2}, the standard deviations quantify the noises of the marginal statistics obtained by the simultaneously readings of the detectors of the measuring device. 

The contribution of the Arthurs-Kelly has triggered a wide variety of research, as the conceptual understanding of the simultaneous measurements of non-commuting observables \cite{Arthurs1965, She1966, Park1968, Stenholm1992, Heinosaari2016}, generalizations to unbiased simultaneous measurements for any non-commuting variables \cite{Arthurs1988, Ishikawa1991, Ozawa1991}, simultaneous measurements in discrete variable systems \cite{Allahverdyan2010, *Ruskov2010, *Garcia2016, *Hacohen2016, *Perarnau2017, *Chantasri2018}, weak measurements \cite{Piacentini2016, Ochoa2018}, precision enhancement with correlated detectors \cite{Bullock2014}, accuracy, error and disturbance concepts \cite{Appleby1998, *Appleby1998a, *Appleby1998b, *Appleby1998c, Mendoza2021}, up to the potential field of quantum information, as entanglement generation, teleportation, entanglement swapping, and transfer information through quantum tomography \cite{Roy2014, Sahu2018}.  

In this work, we pose the Arthurs-Kelly model as a generator of CV tripartite entanglement in the specific regime of Gaussian states. Although the original proposal entails the measurement process on an arbitrary state, we appeal to a particular Gaussian state as the system under measurement mainly to theoretical and practical advantages. For example, they are the common referent in quantum communication and teleportation tasks \cite{Furusawa1998, *Braunstein1998, *Milburn1999, *Bowen2003, *Yonezawa2004, *Ban2004, *Adesso2005f, *Takei2005, *Hammerer2005, *Adesso2007, *Zhang2009, *Smith2011, *Srikara2020} or in quantum key distribution and quantum cryptography \cite{Hillery2000, *Cerf2001, *Grosshans2002, *Silberhorn2002, *Gottesman2001, *Grosshans2003, *Weedbrook2004, *Rodo2007, *Zhou2018}. Besides, the Gaussian states have a well-developed theoretical background for entanglement detection and entanglement measures \cite{Giedke2001, Adesso2006,  Adesso2006a}, which allows a straightforward evaluation of the entanglement properties in our particular measurement scheme. 

\subsection{\label{subsec:l} Measurement configuration}

The original measurement scheme consider a couple of pointer detectors which are represented by balanced centred Gaussian states with finite squeezing \cite{Menicucci2006}, that is,
\begin{equation}
\left|0, V_{j}\right\rangle_{x}=\left(\pi V_{j}\right)^{-\frac{1}{4}} \int dx_{j}~e^{-x_{j}^2/2V_{j}} \left|x_{j}\right\rangle,~~j=1,2,\label{eq:3}
\end{equation}
with $V_{j} < 1$; therefore, being both detectors squeezed in $x_{j}$ direction.  The label $x$ remind us that we define the state in the position space through the superposition of the position quadrature basis $\left\lbrace \left|x_{j}\right\rangle \right\rbrace_{x_j \in \mathbb{R}}$ with the Gaussian wave packet $\phi \left(x_{j}\right)=\left(\pi V_{j} \right)^{-1/4}e^{-x_{j}^{2}/2V_{j}}$, being $\delta_{\hat{x}_{j}}^2= V_{j}/2$ its variance. By a Fourier transform, the representation in momentum space is
\begin{equation}
\left|0, V_{j}\right\rangle_{p}=\left(\frac{V_{j}}{4\pi}\right)^{\frac{1}{4}} \int dp_{j}~e^{-V_{j} p_{j}^{2}/8}\left|p_{j}\right\rangle,~~j=1,2,\label{eq:4}
\end{equation}
expanded in the momentum quadrature basis $\left\lbrace \left|p_{j}\right\rangle \right\rbrace_{p_{j}\in \mathbb{R}}$ and variance  $\delta_{\hat{p}_{j}}^2= 2/V_{j}$. Therefore, the quadrature basis of the states, Eqs. \eqref{eq:3} and \eqref{eq:4}, are connected through
\begin{equation}
\left|q\right\rangle=\frac{1}{2\sqrt{\pi}}\int dp~e^{-iqp/2} \left|p\right\rangle,\label{eq:5}
\end{equation}
\begin{equation}
\left|p\right\rangle=\frac{1}{2\sqrt{\pi}}\int dp~e^{iqp/2} \left|q\right\rangle. \label{eq:6}
\end{equation}
For the Arthurs-Kelly pointers we have $V_{1}=b$, $V_{2}=1/b$, hence the states, Eq. \eqref{eq:3} and \eqref{eq:4}, are related via $V_{1}=\left(V_{2} \right)^{-1}$. The so-called balance parameter $b$ allows to handle the measurement accuracy for any of the two conjugate variables \cite{Busshardt2010}. Moreover, through the states given by Eqs. \eqref{eq:3} and \eqref{eq:4}, it is easy to verify the saturation of the Heisenberg uncertainty relation, i.e., $\delta_{\hat{x}_{j}}^2 \delta _{\hat{p}_{j}}^2 = 1$; therefore, they are minimum uncertainty states, which is a necessary characteristic to reduce the inaccuracy affecting the statistics of the measurement outputs, since the detectors contribute with its own noises to the measurement process \cite{Stenholm1992, Appleby1998, *Appleby1998a, *Appleby1998b, *Appleby1998c}.

For the system under measurement, we choose the most general one-mode pure Gaussian state, that is, a rotated, displaced, and squeezed vacuum state,
$\left|\alpha,\theta, r\right\rangle$, which is defined in terms of the vacuum state as \cite{Weedbrook2012}
\begin{equation}
\left|\alpha,\theta, r\right\rangle=\hat{D}(\alpha) \hat{R}(\theta) \hat{S}(r)\left|0\right\rangle, \label{eq:7}
\end{equation}
with the unitary operators
\begin{equation}
\hat{D}(\alpha) \equiv \exp\left[\alpha \hat{a}^{\dagger} - \alpha^{\ast}\hat{a} \right], \label{eq:8}
\end{equation}
\begin{equation}
\hat{S}(r) \equiv \exp\left[r\left(\hat{a}^2 - \hat{a}^{\dagger 2}\right)/2 \right], ~~r\in\mathbb{R},\label{eq:9}
\end{equation}
\begin{equation}
\hat{R}\left(\theta \right) \equiv \exp\left[-i\theta \hat{a}^{\dagger}\hat{a} \right],~~0\leq \theta \leq 2\pi, \label{eq:10}
\end{equation}
being $\hat{D}(\alpha)$, $\hat{S}(r)$ and $\hat{R}\left(\theta \right)$ the displacement, the one-mode squeezing, and the rotation operators respectively. The squared modulus of the complex amplitude $\alpha=(q + ip/2)$ is related to the average energy of the single-mode system, the squeezing parameter $r$ governs its degree of squeezing, and the angle $\theta$ represents a phase displacement with respect to a local oscillator. Using the operators, Eqs. \eqref{eq:8} to \eqref{eq:10}, the quadratures in the Heisenberg picture for the state $\left|\alpha,\theta, r\right\rangle$ are 
\begin{equation}
\hat{q}_{3}=\left(\hat{q} + q\right)e^{-r}\cos \theta + \left(\hat{p} + p \right)e^{r}\sin \theta,  \label{eq:11}
\end{equation}
\begin{equation}
\hat{p}_{3}=-\left(\hat{q} + q\right)e^{-r}\sin\theta + \left(\hat{p} + p \right)e^{r}\cos \theta,\label{eq:12}
\end{equation}
where $\hat{q}$ and $\hat{p} $ is the initial canonical pair of the vacuum. It is straightforward verify the mean values 
\begin{equation}
\left\langle \hat{q}_{3} \right\rangle=qe^{-r}\cos \theta + pe^{r}\sin \theta,\label{eq:13}
\end{equation}
\begin{equation}
\left\langle \hat{p}_{3} \right\rangle=-q e^{-r}\sin\theta + pe^{r}\cos \theta.\label{eq:14}
\end{equation}
and the variances
\begin{equation}
\delta_{\hat{q}_{3}}^2 = \frac{\left(\cos \theta \right)^2}{e^{2r}} +  e^{2r} \left(\sin \theta \right)^2, \label{eq:15}
\end{equation}
\begin{equation}
\delta_{\hat{p}_{3}}^2 
=\frac{\left(\sin \theta \right)^2}{e^{2r}} +  e^{2r} \left(\cos \theta \right)^2.\label{eq:16}
\end{equation}
Then, within this considerations, the system under measurement does not necessarily represent a minimum uncertainty state, unless that $r=0$ or $\theta=n \pi/2,~n=0, 1, 2, \cdots$; in this specific case, the system under measurement is given by a symmetric Gaussian quasi-probability distribution (Wigner function) in the phase space. 
Hence we have the initial state of the measurement setting as the product
\begin{equation}
\left|\psi\right\rangle_{t=0}=\left|0, V_{1}\right\rangle_{x, p}\left|0, V_{2}\right\rangle_{x, p}\left|\alpha,\theta, r\right\rangle, \label{eq:17}
\end{equation} 
where the label $x,p$ in the state of the detectors means that we can express them in the position or momentum representation. Therefore, we take each subsystem as a CV mode with a associated bosonic operator $\hat{a}_{k}, ~k=1,2,3$, and quadratures $\hat{q}_{k}=(\hat{a}_{k}^{\dagger} + \hat{a}_{k})$ and $\hat{p}_{k}=i(\hat{a}_{k}^{\dagger} - \hat{a}_{k})$, just as those of a quantum harmonic oscillator or a single-mode of the quantum electromagnetic field                                                                                                          . Therefore, within this considerations we have an entirely pure CV Gaussian measurement setup.
\subsection{Dynamics}
To obtain the dynamics of the measurement process, we consider the Hamiltonian interaction, Eq. \eqref{eq:1}, to define the time evolution operator as $\exp\left[-it\hat{H}/2\right]$. We identify $\hat{A}=-\frac{it\kappa_{1}}{2}\hat{x}_{3} \hat{p}_{1}$ and $\hat{B}=-\frac{it\kappa_{2}}{2}\hat{p}_{2}\hat{p}_{3}$, hence we have the following commutators
\begin{align}
\begin{split}
\left[\hat{A}, \hat{B} \right]=&-\frac{i t^{2}\kappa_{1} \kappa_{2}}{2} \hat{p}_{1} \hat{p}_{2}, \\
\left[\hat{A},\left[\hat{A}, \hat{B}\right]\right]=&0,\\ \left[\hat{B},\left[\hat{A}, \hat{B}\right]\right]=&0;  \label{eq:18}
\end{split}
\end{align}
then, using the Baker-Campbell-Hausdorff formula: $e^{\hat{A} + \hat{B}}=e^{-\frac{1}{2}\left[\hat{A}, \hat{B} \right]}e^{\hat{A}} e^{\hat{B}}$, we factorize the time evolution operator as
\begin{equation}
\hat{U}(t)=e^{\frac{i\alpha_{1} \alpha_{2}}{4}\hat{p}_{1}\hat{p}_{2}}e^{-\frac{i\alpha_{1}}{2}\hat{x}_{3}\hat{p}_{1}}e^{-\frac{i\alpha_{2}}{2}\hat{p}_{3}\hat{p}_{2}}, \label{eq:19}
\end{equation}
where we take $\alpha_{j}=\kappa_{j}t$ as the \textit{displacement strength}, which displaces the expected value of the position of pointer $j$ by a quantity proportional to $\kappa_{j}$. Using the unitary operator, Eq. \eqref{eq:19}, the dynamics of the set of observables $\hat{X}'=\left\lbrace \hat{x}_{1}', \hat{x}_{2}', \hat{x}_{3}', \hat{p}_{1}', \hat{p}_{2}', \hat{p}_{3}'  \right\rbrace$ in the Heisenberg picture is
\begin{align}
\begin{split}
\hat{x}_{1}'=&\hat{x}_{1} + \alpha_{1}\hat{x}_{3} +\frac{\alpha_{1}\alpha_{2}}{2}\hat{p}_{2}, \\
\hat{x}_{2}'=&\hat{x}_{2} + \alpha_{2}\hat{p}_{3} -\frac{\alpha_{1}\alpha_{2}}{2}\hat{p}_{1}, \\
\hat{x}_{3}'=&\hat{x}_{3} + \alpha_{2} \hat{p}_{2}, \\
\hat{p}_{1}'=&\hat{p}_{1}, \\
\hat{p}_{2}'=&\hat{p}_{2}, \\
\hat{p}_{3}'=&\hat{p}_{3}- \alpha_{1}\hat{p}_{1}.
\end{split}\label{eq:20}
\end{align}
Where $\hat{X}'=\hat{U}^{\dagger}(t) \hat{X} \hat{U}(t)$ represent the dynamics of the observable $\hat{X}\in\left\lbrace \hat{x}_{1}, \hat{x}_{2}, \hat{x}_{3}, \hat{p}_{1}, \hat{p}_{2}, \hat{p}_{3}  \right\rbrace$ in the Heisenberg scenario; we omit the time dependence for brevity. From Eqs. \eqref{eq:20} it is verified that the dynamics of the global system depend on the intensity strengths $\alpha_{i}$. 
Such quantities depend directly on the couplings $\kappa_{j}$, which acquire meaning according to the physical scenario in which the simultaneous measurement process is carried out. For example, in a non-linear optical context \cite{Stenholm1992}, where two optical signals (acting as the pointers) interact with another (which we pretend to measure) in a non-linear region pumped by a strong field, the coupling is determined by the average energy of the pumped field.
Or in the regime of matter-light interaction \cite{Power1997}, where a free atom interacts with two orthogonal cavity fields, causing them phase displacements proportional to its canonical pair. In this scheme, the couplings depend proportionally on the dipolar coupling between the atom and field and depend reciprocally proportional to the atom-field detuning. However, it is important to note that the displacement strengths can take any value at the judgment of the experimentalist by simply adjusting the measurement scale of the pointers \cite{Stenholm1992}. This is particularly important since the simultaneous measurement schemes for non-commuting observables are mathematically based on the joint unbiasedness condition, which demands that the marginal mean values obtained through the measurement apparatus match with the theoretical values of the observables under inspection \cite{Arthurs1965, Arthurs1988, Ozawa1991, Ishikawa1991}. From the dynamics, Eqs.  \eqref{eq:12}, we can verify that
\begin{equation}
\left\langle \hat{x}_{1}'\right\rangle = \alpha_{1} \left\langle \hat{x}_{3}\right\rangle, \label{eq:21}
\end{equation}
\begin{equation}
\left\langle \hat{x}_{2}'\right\rangle = \alpha_{2} \left\langle \hat{p}_{3}\right\rangle; \label{eq:22}
\end{equation}
then, for equal measurement scales of both detectors (i.e., same coupling constants $\kappa_{1}=\kappa_{2}=\kappa$), it is required the time of measurement as $\tau=(\kappa)^{-1}$; however, at the time $\tau'=(\kappa_{i})^{-1}$ and for distinct coupling constants $\kappa_{i} \neq \kappa_{j}$, it is necessary to adjust the measurement scale of pointer $j$ at the rate $\kappa_{j}/\kappa_{i}$ to still holding the joint unbiasedness condition. In the subsequent we follow to Arthurs and Kelly taking $\alpha_{1}=\alpha_{2}=1$.

On the other hand, Arthurs and Kelly showed that the marginal statistics of the measured position and momentum observables are displaced from their theoretical values as a consequence of the inherent noise carried by the measurement process, which is proportional to the fluctuations of the detectors, i.e., to the balance parameter; using the dynamics, Eq. \eqref{eq:20}, and the states of the detectors, Eqs. \eqref{eq:3} and \eqref{eq:4}, we mean
\begin{equation}
\delta_{\hat{x}_{1}'}^2=\delta_{\hat{x}_{3}}^2  + \eta_{1}\left(b\right), \label{eq:23}
\end{equation} 
\begin{equation}
\delta_{\hat{x}_{2}'}^2=\delta_{\hat{p}_{3}}^2+ \eta_{2}\left(b\right),\label{eq:24}
\end{equation}
where $\eta_{1}\left(b\right)=\delta_{\hat{x}_{1}}^2 + \frac{\delta_{\hat{p}_{2}}^2}{4}=b$, and $\eta_{2}\left(b\right)=\delta_{\hat{x}_{2}}^2 + \frac{\delta_{\hat{p}_{1}}^2}{4}=\frac{1}{b}$. It is straightforward to verify that the product $\delta_{\hat{x}_{1}'}^2 \delta_{\hat{x}_{2}'}^2$ is minimized by the balance parameter adjusted at the rate $b=\frac{\delta_{\hat{x}_{3}}}{\delta_{\hat{p}_{3}}}$ for \textit{any} quantum state as the system under measurement; then, in the follows, we use such optimal value to maintain the fidelity with the Arthurs-Kelly proposal.

\begin{figure}
\includegraphics[width=0.48\textwidth]{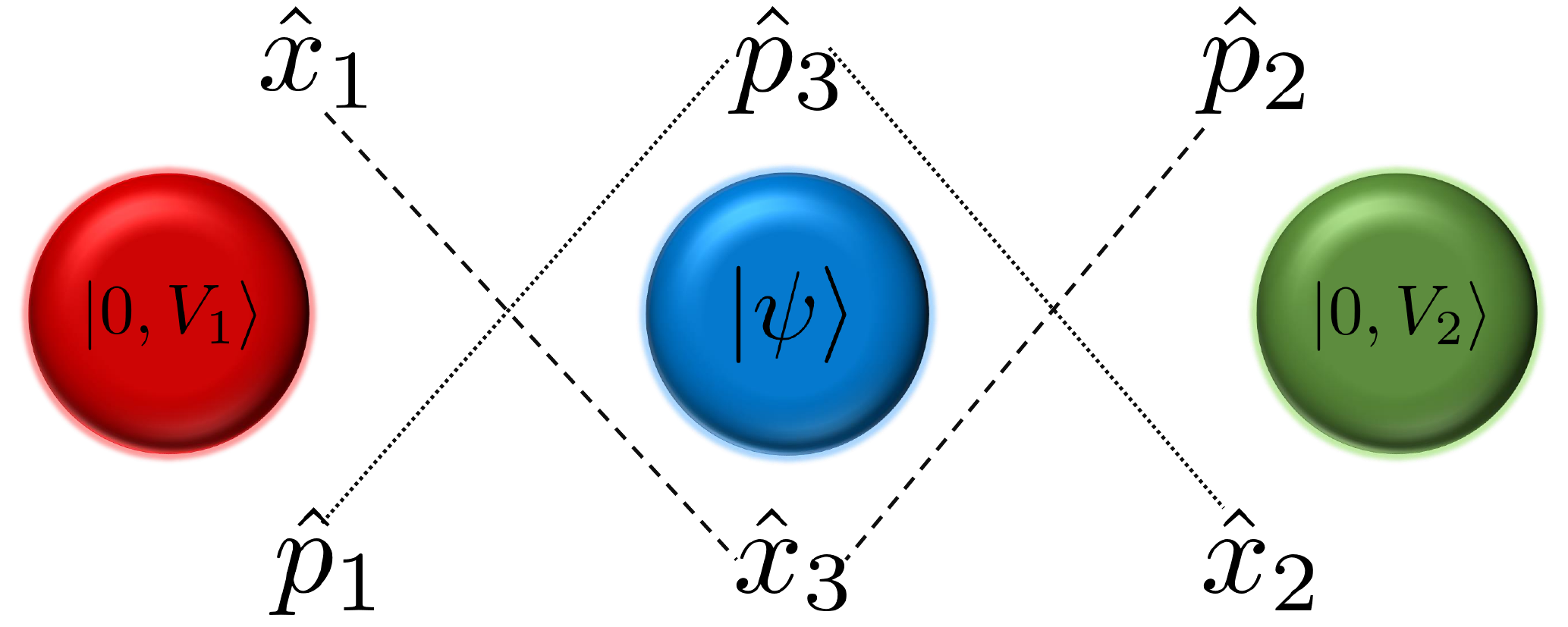}
\centering
\caption{Tripartite entanglement implied by the Arthurs-Kelly measurement process. The dynamics (see Eq. \eqref{eq:20}) of $x_{1}'$ and $\hat{x}_{3}'$ shows an tripartite correlation among the $\hat{x}_{1}, \hat{x}_{3}$ and $\hat{p}_{2}$ variables (dashed lines). The dynamics of $x_{2}'$ and $\hat{p}_{3}'$ shows an tripartite correlation among the $\hat{p}_{1}, \hat{x}_{2}$ and $\hat{p}_{3}$ variables (dotted lines). The red circle represents the first detector, the green one the second detector, and the blue circle constitutes the system under inspection, in this case a Gaussian state.}\label{fig:1}
\end{figure}

The dynamics displayed by Eqs. \eqref{eq:20} suggests the existence of quantum correlations among the three CV modes constituting the measurement setting. By inspecting the $\hat{x}_{1}'$ and $\hat{x}_{3}'$ variables we observe correlations among $\hat{x}_{1}, \hat{x}_{3}$ and $\hat{p}_{2}$, and through  $\hat{x}_{2}'$ and $\hat{p}_{3}'$, we observe correlations between $\hat{p}_{1}, \hat{x}_{2}$ and $\hat{p}_{3}$; then, both relations implying a tripartite entanglement relation between three distinct observables of the measurement configuration; see Fig. \ref{fig:1}. In the subsequent we will focus on to prove the existence of genuine tripartite entanglement in the measurement configuration as Eqs. \eqref{eq:20} suggest.


\section{Gaussian states and tripartite continuous variable entanglement} \label{Sec. 3}
\subsection{\label{sec2A} Continuous variable Gaussian states}
With the considerations of the last section we have a  CV system with three entangled bosonic modes; then, the formalism of $N$ bosonic systems in a `nutshell' is valid \cite{Weedbrook2012}. In general, each mode has an associated Hilbert space $\mathcal{H}_{k}$ and a pair of canonical variables of position and momentum $\hat{x}_{k}'$ and $\hat{p}_{k}'$ which we array in a quadrature vector: $\hat{\textbf{R}}=\left(\hat{x}_{1}', \hat{p}_{1}', \cdots, \hat{x}_{N}', \hat{p}_{N}'  \right)^T$. The commutation relation between the $i$ and $j$ component of $\hat{\textbf{R}}$ satisfy $\left[\hat{x}_{i}', \hat{x}_{j}' \right]=2i \Omega_{ij},$ with $\Omega_{ij}$ the generic element of the following symplectic matrix
\begin{equation}
\boldsymbol \Omega= \bigoplus_{k=1}^{N} \boldsymbol 
w = \left(
\begin{array}{ccc}
\boldsymbol w & & \\
& \ddots & \\
& & \boldsymbol w
\end{array}\right),  ~ \boldsymbol w= \left(
\begin{array}{cc}
0&1  \\
-1&0
\end{array}\right). \label{eq:25}
\end{equation}
In particular, the Gaussian states are a referent from the quantum optics domain; the family includes the vacuum, coherent, thermal, and squeezed states. By definition they are states whose characteristic,  $\chi(\boldsymbol \xi)$, and Wigner function, $W (\hat{\textbf{R}})$, is Gaussian \cite{Braunstein2005, Weedbrook2012}; where these functions are defined as
\begin{equation}
\chi(\boldsymbol \xi)=\text{Tr}\left[\hat{\rho} \hat{D}(\boldsymbol \xi) \right],\label{eq:26a}
\end{equation} 
\begin{equation}
W (\hat{\boldsymbol R})=\int \frac{d^{{2N}} \boldsymbol \xi}{(2\pi)^{2N}} \exp \left(-i\hat{\boldsymbol R}^{T} \boldsymbol \Omega \boldsymbol \xi \right) \chi(\boldsymbol \xi), \label{eq:26b}
\end{equation}
with $\hat{D}(\boldsymbol \xi) = \exp\left[i \hat{\boldsymbol R}^T \boldsymbol \Omega \boldsymbol \xi \right]$ being the $N$-mode displacement operator and $\boldsymbol \xi \in \mathbb{R}^{2N}$ a vector belonging to the $2N$-dimensional phase space of the system.

 The Gaussian states have the goodness of being described only by the first and second moments. The first moments  are made up by the vector of expected values 
\begin{equation}
\overline{\boldsymbol R} = \left\langle \hat{\boldsymbol R} \right\rangle=\text{Tr}\left[\hat{\rho}\hat{\boldsymbol R} \right]; \label{26}
\end{equation}
while the second moments build up the $(2N \times 2N)$-dimensional covariance matrix (CM) $\boldsymbol \sigma$, whose arbitrary $ij$-element is defined as
\begin{equation}
\sigma_{ij}=\frac{1}{2} \left\langle \left\lbrace \Delta \hat{R}_{i}, \Delta \hat{R}_{j} \right\rbrace \right\rangle,\label{eq:27}
\end{equation}
with $\Delta \hat{R}_{i}=\hat{R}_{i} - \left\langle \hat{R}_{i} \right\rangle$ and $\left\lbrace \cdot \right\rbrace$ representing the anticommutator. The CM has the properties to be a real, symmetric and positive definite matrix, which contains all classical and quantum information between the modes conforming the system; furthermore, it contains all locally invariant information as entanglement and purity \cite{Adesso2006}. 

Then, equipped with the last definitions, the characteristic and Wigner functions for Gaussian states are
\begin{equation}
\chi(\boldsymbol \xi)= \exp\left[-\frac{1}{2}\boldsymbol \xi^{T}\left(\boldsymbol {\Omega \sigma \Omega}^{T} \right) \boldsymbol \xi - i (\boldsymbol{\Omega \overline{R}})^{T} \boldsymbol \xi \right], \label{27a}
\end{equation}
\begin{equation}
W \left(\boldsymbol R \right)= \frac{\exp\left[-\frac{1}{2}\left(\boldsymbol {R - \overline{R}}  \right)^T \boldsymbol \sigma^{-1}\left(\boldsymbol {R - \overline{R}}  \right) \right]}{(2\pi)^{N}\sqrt{\text{det}~\boldsymbol \sigma}}. \label{27b}
\end{equation}
On the other hand, the physical validity of a CM is only guaranteed if it satisfy the Heisenberg uncertainty relation in terms of the symplectic form \cite{Serafini2017}
\begin{equation}
\boldsymbol \sigma + i \boldsymbol \Omega \geq 0.\label{eq:28}
\end{equation}
Gaussian unitary operations are those which map Gaussian states to Gaussian states. Gaussian unitary operations affecting only the first moments come through first-order polynomials in quadrature operators, just as the displacement operator; notably, they do not have an impact on the entanglement properties of the system since they can be adjusted locally \cite{Adesso2006, Adesso2014}. Gaussian unitary operations in Hilbert space coming from the second-order polynomials in quadrature operators are equivalent to symplectic transformations in phase space; they act on the first and second moments as: $ \hat{\boldsymbol R}'=\boldsymbol {S \hat{R}},~\boldsymbol \sigma'={\boldsymbol {S  \sigma  S}}^T$, where the matrix $\boldsymbol S$ belongs to the so-called symplectic group, $\text{Sp}(2N, \mathbb{R})$, spanned by the set of $(2N \times 2N)$-dimensional invertible matrices that satisfy $\boldsymbol {S \Omega S}^T=\boldsymbol \Omega$. For further details on the symplectic analysis in quantum mechanics, see \cite{Dutta1995, De2006}.

Using the symplectic framework, there exist a symplectic transformation $\boldsymbol S $ such that $\boldsymbol S^T \boldsymbol {\sigma S} = \boldsymbol \nu$, with
\begin{equation}
\boldsymbol \nu= \oplus_{j=1}^{N} \text{diag}\left(\nu_{j}, \nu_{j} \right), \label{eq:29}
\end{equation}
where the set $\left\lbrace \nu_{j} \right\rbrace$ are the symplectic eigenvalues of $\boldsymbol \sigma$, while $\boldsymbol \nu$ is referred as the Williamson normal form of $\boldsymbol \sigma$ \cite{Williamson1936}. The set $\left\lbrace \nu_{j} \right\rbrace$ can be obtained through the eigenvalues of the matrix $\left|i \boldsymbol \Omega \boldsymbol \sigma \right|$ \cite{Adesso2004, Adesso2005}. Besides, the $ \nu_{j}$ are invariant under symplectic operations on the CM; then, the set codifies essential information about the entanglement of the Gaussian system. In particular, the uncertainty relation, Eq. \eqref{eq:28}, is equivalent to the condition \cite{Adesso2006a}
\begin{equation}
\nu_{j} \geq 1,~~j=1, \cdots, N, \label{eq:30}
\end{equation}
providing an useful method to verify the physical validity of a CM. 

Notably, all marginal and global entanglement properties of a CV tripartite Gaussian system are determined entirely by the CM of the system. In the following sections, we use this fact to study the entanglement generated in the simultaneous measurement process posed by Arthurs and Kelly for our specific Gaussian configuration.

\subsection{\label{subsec:2.2} Tripartite entanglement and the PPT criterion}
\begin{figure*}
\includegraphics[width=0.98\textwidth]{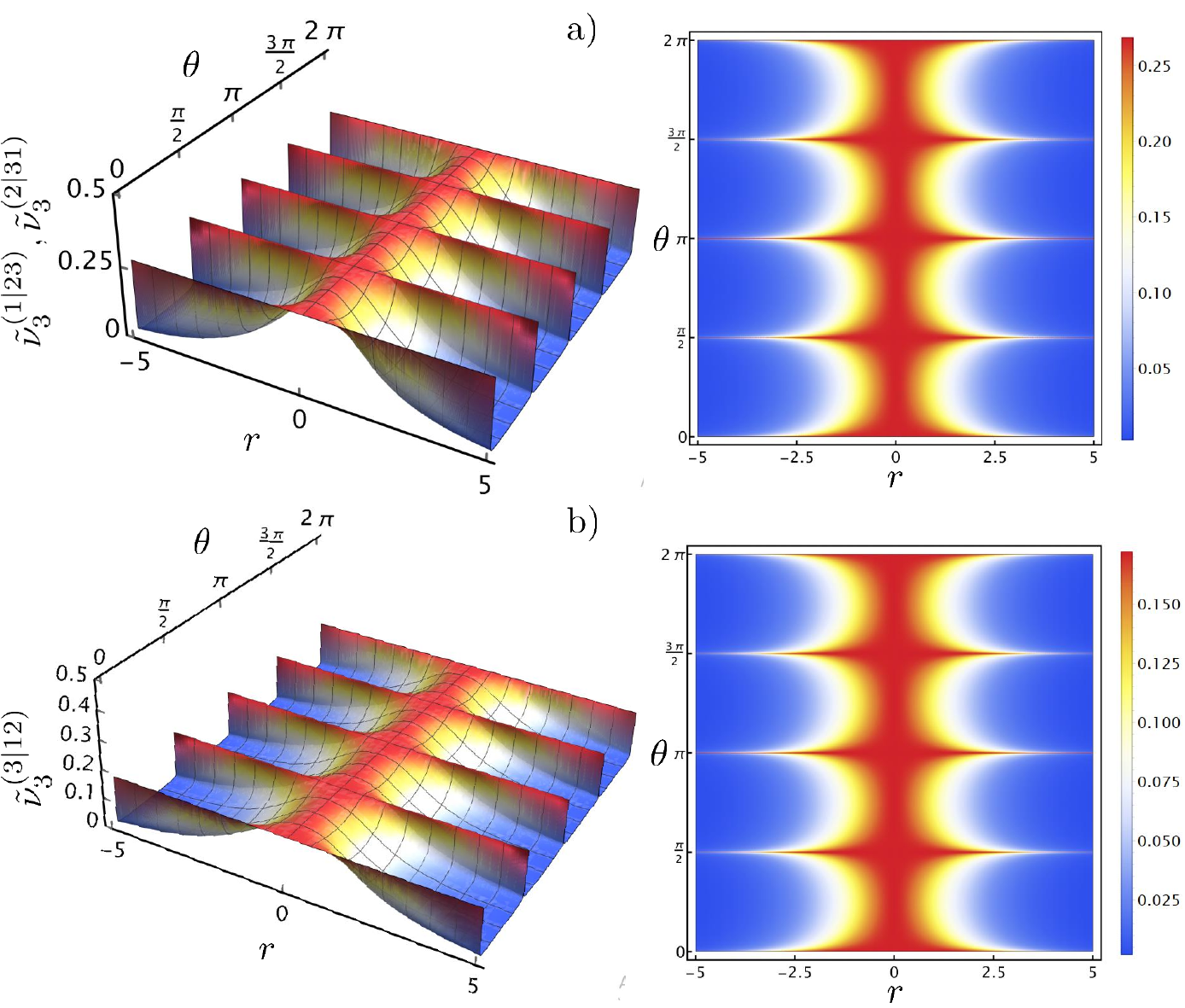}
\caption{\label{fig:2}Plots (left figures) and density plots (right figures) for the symplectic eigenvalues: a) $\tilde{\nu}_{3}^{ \left\lbrace(1|23), (2|31)  \right\rbrace}$  and b) $\tilde{\nu}_{3}^{(3|12) }$ as functions of the squeeze parameter $r$ and the rotation angle $\theta$ of the Gaussian system under measurement. Values less than 1 for these quantities confirm the unphysical validity of the three partially transposed CMs given in Eqs. \eqref{eq:34a} to \eqref{eq:34c}; hence this neglect the corresponding bipartitions, therefore, this implies genuine tripartite entanglement in the measurement setting of Arthurs and Kelly when the system under observation is the most general pure single-mode Gaussian state, that is, a rotated, displaced, and squeezed vacuum state.}
\end{figure*}
Let us first recover one condition to generally establish genuine tripartite entanglement; this goes as follows: three parties, each containing a single system, are genuinely tripartite entangled if the density operator of the whole system cannot be written as the sum of the three  mixtures of pure \textit{biseparable} products of the whole system, that is, in the biseparable form \cite{Hyllus2006, Jungnitsch2011, Teh2014}:
\begin{equation}
\hat{\rho}= P_{1} \sum_{i} p_{i}\hat{\rho}_{1}^{i} \hat{\rho}_{23}^{i} +  P_{2} \sum_{j} p_{j}\hat{\rho}_{2}^{j} \hat{\rho}_{13}^{j} +  P_{3} \sum_{k} p_{k}\hat{\rho}_{3}^{k} \hat{\rho}_{12}^{k}, \label{eq:30.1}
\end{equation}
where $P_{m}$ and $p_{n}$ are normalized probability distributions, then $\sum_{m=1}^{3} P_{m}=1$ and $\sum_{n} p_{n}=1$. Besides, $\hat{\rho}_{j}^{i}$ is the density operator representing the single-mode system $j$, while $\hat{\rho}_{kl}^{i}$ is the density operator representing the joint quantum state of the systems $k$ and $l$, with $j,k,l \in \left\lbrace 1,2,3 \right\rbrace,~ j\neq k\neq l$. Then, each biseparable product $\hat{\rho}_{j}^{i} \hat{\rho}_{kl}^{i}$ implies the systems $k$ and $l$ are entangled, but there is no entanglement relation between $j$ with $k$ or $j$ with $l$. Therefore, by neglecting the three convex sums of biseparable states of a three-mode state we confirm genuine tripartite entanglement. 

By focusing on the regime of pure states, the general condition, Eq. \eqref{eq:30.1}, implies that by neglecting the three possible bipartitions of the tripartite system, we confirm genuine tripartite entanglement. This is an interesting fact since, for any system composed by $N$ parties, the positivity of its partially transposed density matrix is a necessary condition for its separability with respect to any bipartition \cite{Peres1996}; this statute is named as the positivity of partial transposition (PPT) criterion. For any CV quantum state, the action of partial transposition of its density matrix is reflected on the CM as a sign flip in the momenta of the system with respect to which is transposing \cite{Simon2000}; that is, considering a general bipartition $A|B$ of a system of $(m+n)$-modes, such that the parts $A$ and $B$ comprise $m$ and $n$ systems respectively, the partial transposition respect to $A$ is obtained through
\begin{equation}
\tilde{\boldsymbol \sigma}_{m|n}=\textbf{T} \boldsymbol \sigma_{m|n} \textbf{T},~~\textbf{T}=\oplus_{1}^{m} \left(
\begin{array}{cc}
1&0  \\
0&-1
\end{array}\right) \oplus \mathbb{I}_{2n\times 2n},\label{eq:31}
\end{equation}
where $\mathbb{I}_{2n\times 2n}$ represent the $(2n\times 2n)$-dimensional identity matrix. Remarkably, the PPT criterion is a condition necessary and sufficient for the separability of (1 vs $N-1)$ modes in CV Gaussian states \cite{Simon2000, Werner2001}. Besides, it is important to note that the $(i|jk)$-mode bipartitions encompass all possible bipartite splits in a three-mode Gaussian state; therefore, by neglecting them, we will prove genuine entanglement among the three parties; see Ref. \cite{Hyllus2006}. Then, for a determined bipartition, the separability will be only guaranteed if the corresponding partially transposed CM satisfies the Eq. \eqref{eq:8}; therefore, $\tilde{\boldsymbol \sigma}_{i|jk} + i \boldsymbol \Omega \geq 0$; or, analogously, if its symplectic eigenspectrum satisfies  $\tilde{\nu}_{i|j k} \geq 1$ as Eq. \eqref{eq:30} establishes. It must be noted that we use indistinctly the terms non-separability and entanglement, which essentially imply distinct concepts \cite{Hyllus2006}; however, let us remember that we are tackling a problem that involves only pure states; therefore, in that context, both concepts converge \cite{Teh2014}. In the following, we use the PPT criterion to determine the separability of the Arthurs-Kelly measurement process for our specific Gaussian configuration. In all the following, we omit the $(r, \theta)$-dependence in the pertinent quantities for brevity.

Using the dynamics exposed by Eqs. \eqref{eq:20}, we define the following three quadrature vectors
\begin{equation}
\hat{\textbf{R}}_{1}= \left(\hat{\textbf{C}}_{1},\hat{\textbf{C}}_{2}, \hat{\textbf{C}}_{3} \right)^T, \label{eq:32}
\end{equation}
\begin{equation}
\hat{\textbf{R}}_{2}= \left(\hat{\textbf{C}}_{3},\hat{\textbf{C}}_{1}, \hat{\textbf{C}}_{2} \right)^T, \label{eq:33}
\end{equation}
\begin{equation}
\hat{\textbf{R}}_{3}=\left(\hat{\textbf{C}}_{2},\hat{\textbf{C}}_{3}, \hat{\textbf{C}}_{1} \right)^T, \label{eq:34}
\end{equation}
with $\hat{\textbf{C}}_{i} =\left(\hat{x}_{i}', \hat{p}_{i}' \right);$ then, by employing the definition, Eq. \eqref{eq:27}, we obtain the following three $(6 \times 6)$-dimensional CMs in block form 
\begin{equation}
\boldsymbol \sigma_{1|23}=  \begin{pmatrix}
    \begin{array}{ccc}
 \boldsymbol \sigma_{1}  & \boldsymbol \varepsilon_{1, 2} &\boldsymbol \varepsilon_{1, 3}  \\
\boldsymbol \varepsilon_{1, 2}^T & \boldsymbol \sigma_{2} & \boldsymbol \varepsilon_{2, 3} \\
\boldsymbol \varepsilon_{1, 3}^T& \boldsymbol \varepsilon_{2,3}^T & \boldsymbol \sigma_{3}
    \end{array}
  \end{pmatrix}, \label{eq:34a}
\end{equation}
\begin{equation}
\boldsymbol \sigma_{3|12}=  \begin{pmatrix}
    \begin{array}{ccc}
  \boldsymbol \sigma_{3}  & \boldsymbol \varepsilon_{1, 3}^T & \boldsymbol \varepsilon_{2, 3}^T  \\
\boldsymbol\varepsilon_{1, 3} & \boldsymbol \sigma_{1} & \boldsymbol \varepsilon_{1, 2} \\
\boldsymbol \varepsilon_{2, 3} & \boldsymbol\varepsilon_{1, 2}^T & \boldsymbol \sigma_{2}
    \end{array}
  \end{pmatrix}, \label{eq:34b}
\end{equation}
\begin{equation}
\boldsymbol \sigma_{2|31}=  \begin{pmatrix}
    \begin{array}{ccc}
  \boldsymbol \sigma_{2}  & \boldsymbol\varepsilon_{2, 3} & \boldsymbol\varepsilon_{1, 2}^T  \\
\boldsymbol\varepsilon_{2, 3}^T & \boldsymbol\sigma_{3} & \boldsymbol\varepsilon_{1, 3}^T \\
\boldsymbol\varepsilon_{1, 2}& \boldsymbol\varepsilon_{1, 3} & \boldsymbol\sigma_{1}
    \end{array}
  \end{pmatrix}, \label{eq:34c}
\end{equation}
where the $\boldsymbol \sigma_{i}$ and $\boldsymbol \varepsilon_{j, k}$ elements are $(2 \times 2)$-dimensional matrices defined in the Appendix \ref{appendixA}. It can be verified that the above matrices satisfy the following conditions for pure three-mode CV Gaussian states \cite{Adesso2006a}
\begin{equation}
\text{Det} ~\boldsymbol \sigma =1, \label{eq:35}
\end{equation}
\begin{equation}
\Delta_{1 ,2 ,3}=3, \label{eq:36}
\end{equation}
\begin{equation}
\text{Det} ~\boldsymbol \sigma_{ij} = \text{Det} ~\boldsymbol \sigma_{k},~~i \neq j \neq k, \label{eq:37}
\end{equation}
where 
\begin{equation}
\Delta_{1,2,3} \equiv \sum_{j=1}^{3} \text{Det} ~\boldsymbol \sigma_{j} + 2\sum_{j<k} \text{Det}  ~\boldsymbol \varepsilon_{j,k}, \label{eq:38}
\end{equation}
\begin{figure}
\includegraphics[width=0.48\textwidth]{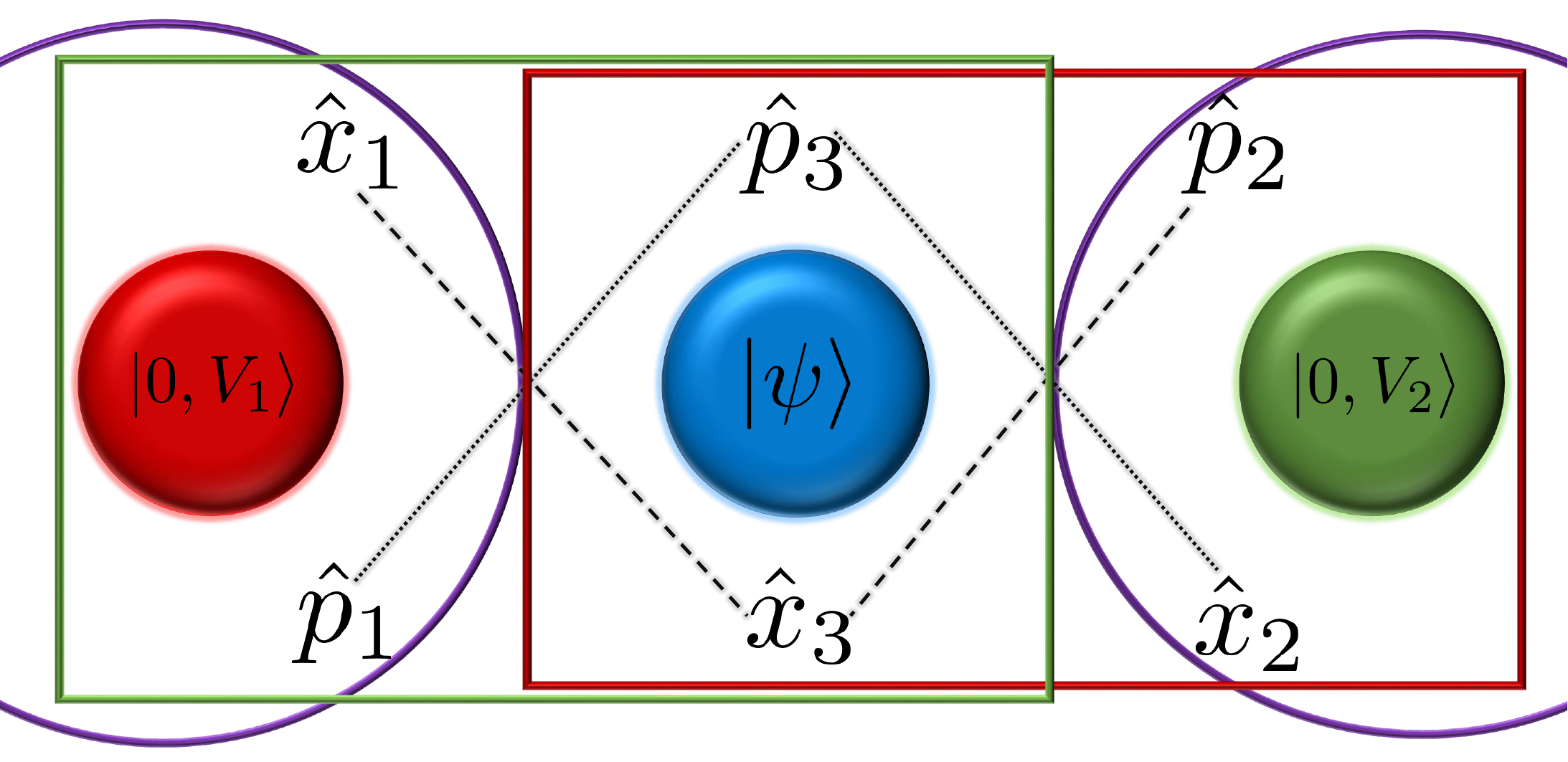}
\centering
\caption{Sketch for the three $i|jk$-mode bipartitions in the Arthurs-Kelly measurement setting. The green rectangle assumes entanglement between the first detector and the Gaussian system but not from them with the second detector (bipartition $2|3 1$). The brown rectangle implies entanglement between the second detector and the Gaussian system but not from them with the first detector (bipartition $1|2 3$). The purple semicircles encompass entanglement among the two detectors but not with the system under measurement (bipartition $3|1 2$). The three depicted bipartitions are not valid due to the tripartite entanglement relation (dotted and dashed lines) between the three conjugate pairs of observables in the whole system.}\label{fig:3}
\end{figure}
is an invariant quantity under symplectic transformations on the CM \cite{Serafini2006f}. Moreover, $\boldsymbol \sigma_{ij}$ represents the CM for the reduced two-mode state comprising the modes $i$ and $j$, which is obtained by removing the entries of the mode $k$ in $\boldsymbol \sigma$. 

Using the definition, Eq. \eqref{eq:31}, and the software \textit{Mathematica}, we partially transpose the CMs, Eqs. \eqref{eq:34a} to \eqref{eq:34c}; then, we compute the symplectic eigenvalues  $\tilde{\nu}_{l}^{i|jk},~i,j,k,l\in \left\lbrace 1,2,3 \right\rbrace \text{with} ~i\neq j \neq k$, as we explain in Sec. \ref{sec2A}; we attach the notebook as supplementary material \cite{[{ See Supplemental Material at [...] for the process of computation of the symplectic eigenvalues and a numerical exploration of their maximum and minimum inside the region delimited by the Eq. \eqref{eq:38.2}. The results are provided as a \textit{Mathematica} file}]Supmat}. In particular, \textit{Mathematica} does not provide close expressions for the symplectic eigenvalues; instead, these quantities are given in terms of the roots of a cubic polynomial whose coefficients are functions of $r$ and $\theta$; see \cite{Supmat}. Notably, we find

\begin{equation}
\tilde{\nu}_{i}^{(1|23)} = \tilde{\nu}_{i}^{(2|31)},~i=1,2,3, \label{eq:38.1}
\end{equation}
which is an indicative of certain symmetry of entanglement arround the system under measurement; we will analyze this fact in Sec. \ref{subsec:D}.

Now, in the following, we will focus on the rectangular window in the plane $r$ and $\theta$
\begin{equation}
\mathcal{R}=\left\lbrace \left(r, \theta\right) |-5\leq r \leq 5,~0 \leq \theta \leq 2\pi  \right\rbrace,  \label{eq:38.2}
\end{equation}
in order to explore numerically the maximum and minimum values of the computed symplectic eigenvalues inside this region; we infer the following three cases (see Sec. 3 of \cite{Supmat}):
\begin{equation}
\tilde{\nu}_{1}^{\left\lbrace (1|23), (3|12), (2|31)\right\rbrace}> 1,\label{eq:39}
\end{equation}
\begin{equation}
\tilde{\nu}_{2}^{\left\lbrace (1|23), (3|12), (2|31)\right\rbrace} = 1,\label{eq:40}
\end{equation}
\begin{equation}
\tilde{\nu}_{3}^{\left\lbrace (1|23), (3|12), (2|31)\right\rbrace} < 1;\label{eq:41}
\end{equation}
in particular, we find the following maximum values: $\text{max}\left[\tilde{\nu}_{3}^{(1|23)} \right] = \text{max}\left[\tilde{\nu}_{3}^{(2|31)}\right] \approx 0.268$ and $\text{max}\left[\tilde{\nu}_{3}^{(3|12)} \right] \approx 0.171$, obtained when the Gaussian system under measurement is a minimum uncertainty state. In Fig. \ref{fig:2} we show the plots and density plots for the symplectic eigenvalues, $\tilde{\nu}_{3}^{i|jk}~ \forall~ r, \theta \in \mathcal{R}$, of the three transposed CMs, Eqs. \eqref{eq:34a} to \eqref{eq:34c}; the existence of values less than 1 for these quantities is equivalent to the violation of the uncertainty relation, Eq. \eqref{eq:28}, for each partially transposed CM; hence, this negates all $(1$ vs $2)$-mode bipartitions of the system; see Fig. \ref{fig:3}. Therefore, we certify the generation of tripartite entanglement in the unitary dynamics of the simultaneous measurement process of Arthurs and Kelly when the system under observation is the most general one-mode Gaussian state, being this fact independent of the squeezing $r$ and the rotation angle $\theta$ of the Gaussian system under observation. 
\subsection{Qualitative entanglement properties}
The Gaussian tripartite entanglement generated in the dynamics of our particular simultaneous measurement model of Arthurs and Kelly can be qualitatively categorized since there exists a complete classification for three-mode Gaussian states according to the separability of each of its three $i|jk$-mode bipartitions \cite{Giedke2001}:

\begin{quote}
(\textit{C1}) Fully inseparable states, which are inseparable through any bipartition.\\
(\textit{C2}) One-mode biseparable states, being separable under only one bipartition.\\
(\textit{C3}) Two-mode biseparable states, being separable through two bipartitions.\\
(\textit{C4}) Three-mode biseparable states, resulting separable in all three bipartitions, but impossible to write as a convex sum of tripartite products of pure one-mode states.\\
(\textit{C5}) Fully separable states, which are separable in the three bipartitions.
\end{quote}

Then, according to the above classification, we conclude that the Gaussian tripartite entanglement generated in the dynamics of the simultaneous measurement process of Arthurs and Kelly (when the system under measurement is the most general pure one-mode Gaussian state) is categorized in the class \textit{(C1)} of fully inseparable states, placing them together the well-known CV Greenberger-Horne-Zeilinger (GHZ) states \cite{Van2001, Aoki2003}, and the tripartite version of the CV Einstein-Podolsky-Rosen (EPR) states, generated in \cite{Armstrong2012, Armstrong2015}. 

\subsection{Quantitative entanglement properties} \label{subsec:D}
A crucial property for a \textit{bona fide} entanglement measure, among other properties, is the so-called \textit{monogamy}, which constrains the bipartite entanglement distribution between the different splits of a quantum system composed of $N$ parties: $A_{1}, \cdots, A_{N}$, where each $A_{i}$ contains one system only. By choosing a reference partie $A_{1}$, or focus, this property is quantitatively established as \cite{Coffman2000}
\begin{equation}
E^{A_{1}|A_{2} \cdots A_{N}} \geq \sum_{i=2}^{N}E^{A_{1}|A_{i}}, \label{eq:42}
\end{equation}
being $E^{A_{1}|A_{2} \cdots A_{N}}$ and $E^{A_{1}|A_{i}}$ adequate entanglement measures quantifying the entanglement between $A_{1}$ with the system formed by the parties $(A_{2}, \cdots, A_{N})$ and the reduced two-mode systems $A_{1}|A_{i}$ respectively. In the particular scenario of CV tripartite Gaussian states, various monotones obey the monogamy relation, as the CV version of the tangle (\textit{contangle}) \cite{Adesso2006, Adesso2006a}, defined in the general context of mixed states as the convex roof of the squared logarithmic negativity. Also, the inequality, Eq.  \eqref{eq:42}, gives rise to the definition of  \textit{minimum residual contangle} as a quantifier of genuine tripartite entanglement in CV Gaussian states. This measure is defined for the focus mode given by the smallest multiplicative inverse of the local single-mode purity  \cite{Adesso2006, Adesso2006a}. 
On the other hand, also we can define tripartite entanglement and correlation measures based on the  R{\'e}nyi-2 entropy, which has been proved to satisfy the strong subadditivity inequality, besides to be a natural measure of information for any multimode Gaussian state \cite{Adesso2012}. This quantity is directly linked with the purity of the quantum state: $-\ln\left(\hat{\rho}^2 \right)$; then, in general, it stands as a measure of ignorance about the preparation of the state \cite{Adesso2014}. In particular, based on this entropy, it has the \textit{residual tripartite R{\'e}nyi-2 entanglement} \cite{Adesso2014}  as an adequate entanglement measure for pure three-mode Gaussian states; hence we use this monotone to quantify the tripartite entanglement proved in section \ref{subsec:2.2}. 

To start the analysis, it must be noted that we can bring the CM of any pure three-mode Gaussian state through local unitary symplectic operations into the following standard form \cite{Adesso2006a}:
\begin{equation}
\boldsymbol{\sigma}_{sf} =\begin{pmatrix}
 a_{1} & 0& b^{+}_{12} & 0 &  b^{+}_{13} & 0 \\ 
 0 & a_{1} & 0 &  b^{-}_{12} & 0 &  b^{-}_{13} \\ 
  b^{+}_{12}& 0 & a_{2} & 0 &  b^{+}_{23} & 0 \\ 
 0 &  b^{-}_{12} & 0 & a_{2} & 0 &  b^{-}_{23} \\ 
  b^{+}_{13} & 0 &  b^{+}_{23} & 0 & a_{3} & 0 \\ 
 0 &  b^{-}_{13} & 0 &  b^{-}_{23} & 0 & a_{3}
 \end{pmatrix},\label{eq:43}
\end{equation}

with

\begin{widetext}
\begin{equation}
b_{ij}^{\pm} \equiv \frac{\sqrt{\left[\left(a_{i}-a_{j}\right)^2 - \left(a_{k} -1\right)^2 \right]\left[ \left(a_{i}-a_{j}\right)^2 - \left(a_{k} +1\right)^2\right]}\pm \sqrt{\left[\left(a_{i}+a_{j}\right)^2 - \left(a_{k} -1\right)^2 \right]\left[\left(a_{i}+a_{j}\right)^2 - \left(a_{k} +1\right)^2 \right]}}{4 \sqrt{a_{i} a_{j}}}, 
 \label{eq:44}
\end{equation}
\end{widetext}
and where
\begin{equation}
a_{i}\equiv \sqrt{\text{Det} ~\boldsymbol \sigma_{i}}, \label{eq:45}
\end{equation}
are the local symplectic eigenvalues associated with the reduced single-mode $i$ with CM $\boldsymbol \sigma_{i}$, that is, the $(2 \times 2)$-dimensional matrices in the main diagonal of $\boldsymbol \sigma$. Let us recall that unitary symplectic operations on the individual modes of a multipartite system do not affect its informational properties; therefore, the standard form, Eq.  \eqref{eq:43}, maintains invariant the entanglement in the three-mode Gaussian system. On the other hand, the local single-mode symplectic eigenvalues, Eq. \eqref{eq:45}, are in a reciprocal relationship with the purities $\mu_{i}$ of the reduced single-mode states, that is, $a_{i}= \mu_{i}^{-1}$; these quantities are invariant under symplectic operations on the individual modes of the system \cite{Adesso2006, Adesso2006a}. Notably, it has been proved that all quantitative entanglement properties of tripartite Gaussian systems can be determined from this quantities \cite{Adesso2006, Adesso2006a}, what in fact can be guessed from the standard form, Eq. \eqref{eq:43}. Besides, the $a_{i}$ are constrained to vary according to the triangular inequality $\left|a_{j} - a_{k} \right| + 1 \leq a_{i} \leq a_{j} + a_{k} -1$ for all index permutations of $i,j,k$, in order the tripartite system represent a valid physical state \cite{Adesso2006, Adesso2006a}. 

Using Eqs. \eqref{eq:34a} to \eqref{eq:34c}, we identify 
\begin{equation}
a_{1}=\sqrt{2 + 2\sqrt{\cosh^2 (2r) - \cos^2 (2\theta) \sinh^2 (2r)}}, \label{eq:46}
\end{equation}
\begin{equation}
a_{2}=\sqrt{2 + 2\sqrt{\cosh^2 (2r) - \cos^2 (2\theta) \sinh^2 (2r)}}, \label{eq:47}
\end{equation}
\begin{equation}
a_{3}= \sqrt{5 + 4\sqrt{\cosh^2 (2r) - \cos^2 (2\theta) \sinh^2 (2r)}}. \label{eq:48}
\end{equation}
These quantities have an oscillatory behavior with a period of $\pi/2$; besides, they are increasing functions as the magnitude of the squeeze parameter also increases.  Their maximum values are attained at $\theta=(2n + 1)\pi/4,~n=0, 1, 2, \cdots$ for a determined $r \neq 0$. Their absolute minimums are reached when the Gaussian system under measurement is a minimum uncertainty state, that is when $r=0$ for any $\theta$, or when $\theta=n \pi/2$ for any $r$.  From the above, we deduce that the reduced single-mode systems decrease their purity as the magnitude of the squeezing of the Gaussian system under measurement grows, which is indicative that the global tripartite entanglement also increases at increasing $\left|r\right|$.

Now, the residual tripartite R{\'e}nyi-2 entanglement, $\mathcal{E}_{2}^{\left(A_{i}| A_{j}| A_{k}\right)}$, which takes a reference mode $A_{i}$ (focus), is defined as \cite{Adesso2014}:
\begin{equation}
\begin{split}
\mathcal{E}_{2}^{\left(A_{i}| A_{j}| A_{k}\right)}&= \mathcal{E}_{2}^{\left(A_{i}| A_{j} A_{k}\right)} -  \mathcal{E}_{2}^{\left(A_{i}| A_{j} \right)} -  \mathcal{E}_{2}^{\left(A_{i}| A_{k} \right)}\\
&=\frac{1}{2}\ln \left( \frac{a_{i}^2}{ g_{j}g_{k}}  \right),
\end{split} \label{eq:49}
\end{equation}
where $\mathcal{E}_{2}^{\left(A_{i}| A_{j} A_{k} \right)}$ and $\mathcal{E}_{2}^{\left(A_{i}| A_{j} \right)}$ represent the R{\'e}nyi-2 entanglements 
of the global and reduced bipartitions,  $A_{i}| A_{j} A_{k}$  and $A_{i}| A_{j}$, respectively. In particular, we have
\begin{equation}
\mathcal{E}_{2}^{\left(A_{i}| A_{j} \right)}=\frac{1}{2} \ln g_{k}, \label{eq:50}
\end{equation}
where
\begin{equation}
  g_{k}=\begin{cases}
    1, & \text{if $a_{k} \geq \sqrt{a_{i}^2 + a_{j}^2 - 1}$},\\
    \frac{\beta}{8 a_{k}^2}, & \text{if $\alpha_{k}< a_{k} <  \sqrt{a_{i}^2 + a_{j}^2 - 1}$}.\\
    \left(\frac{a_{i}^2 - a_{j}^2}{a_{k}^2 - 1}\right)^2, & \text{if $a_{k} \leq  \alpha_{k}$},
  \end{cases} \label{eq:51}
\end{equation}
with the quantities
\begin{widetext}
\begin{equation}
\alpha_{k}=\sqrt{\frac{2\left(a_{i}^2 + a_{j}^2\right) + \left(a_{i}^2 - a_{j}^2 \right)^2 + \left|a_{i}^2 - a_{j}^2\right|\sqrt{\left(a_{i}^2 - a_{j}^2 \right)^2 + 8\left(a_{i}^2 + a_{j}^2 \right)}}{2 \left(a_{i}^2 + a_{j}^2 \right)}}, \label{eq:52}
\end{equation}
\end{widetext}
\begin{equation}
\begin{split}
\beta =& 2a_{1}^2 + 2a_{2}^2 +2a_{3}^2 + 2a_{1}^2a_{2}^2 +2a_{1}^2a_{3}^2 + 2a_{2}^2a_{3}^2 \nonumber \\
 &-a_{1}^4 -a_{2}^4 - a_{3}^4 - \sqrt{\delta} -1, 
 \end{split}\label{eq:53}
\end{equation}
and
\begin{equation}
\begin{split}
\delta &= \left[\left(a_{1} + a_{2} + a_{3} \right)^2 - 1 \right] \left[\left(a_{1} - a_{2} + a_{3} \right)^2 - 1 \right] \\
& \times\left[\left(a_{1} + a_{2} - a_{3} \right)^2 - 1 \right] \left[\left(a_{1} - a_{2} - a_{3} \right)^2 - 1 \right].\label{eq:54}
\end{split}
\end{equation}

On the other hand, the R{\'e}nyi-2 entanglement
associated with the global bipartition $A_{i}|A_{j}A_{k}$ is given by
\begin{equation}
\begin{split}
\mathcal{E}_{2}^{(A_{i}|A_{j}A_{k})} &=\frac{1}{2}\ln~ \left[\text{det}(\boldsymbol \sigma_{i}) \right] \\
&= \frac{1}{2}\ln ~\left[\text{det}(\boldsymbol \sigma_{jk}) \right],
\end{split} \label{eq:54.1}
\end{equation}
being $\boldsymbol \sigma_{i}$ and $\boldsymbol \sigma_{jk}$ the reduced CMs associated with the systems comprising the single-mode $i$ and the two-mode $jk$.

In our tripartite system, $A_{1}$ represents the partie containing the mode of the first detector, $A_{2}$ contains the mode of the second detector and $A_{3}$ comprises the mode of the Gaussian system under observation. Then, to give a complete quantitative description of the entanglement in the dynamics of the measurement process, we analyze the R{\'e}nyi-2 entanglements of all global and reduced mode bipartitions $\mathcal{E}_{2}^{(A_{i}|A_{j}A_{k})}$ and  $\mathcal{E}_{2}^{(A_{i}|A_{j})}$. We begin with the R{\'e}nyi-2 entanglements  $\mathcal{E}_{2}^{(A_{i}|A_{j})}$ for the two-mode reduced states to examine the one-to-one entanglement relations in our Arthurs-Kelly measurement setting. Then, in the following, we use the invariant symplectic quantities, Eqs.  \eqref{eq:46} to \eqref{eq:48}, describing our tripartite Gaussian system.

First, we carry out a numerical minimization for the difference $a_{3} - \sqrt{a_{1}^{2} + a_{2}^2 - 1}, ~\forall~r,\theta$ obtaining $0$, which implies $ a_{3} \geq \sqrt{a_{1}^{2} + a_{2}^2 - 1}$; therefore, according to Eq. \eqref{eq:51} we conclude that $g_{3}=1$; then, due to Eq. \eqref{eq:50}, the above means that
\begin{equation}
\mathcal{E}_{2}^{\left(A_{1}|A_{2}\right)}=0. \label{eq:55}
\end{equation}
For the reduced entanglements $\mathcal{E}_{2}^{\left(A_{1}|A_{3}\right)}$ and $\mathcal{E}_{2}^{\left(A_{2}|A_{3}\right)}$ it must be noticed from Eqs. \eqref{eq:46} and \eqref{eq:47} that $a_{1}=a_{2}=a$; then, from Eq. \eqref{eq:52}, this implies that $\alpha_{1}=\alpha_{2}=\alpha$;  hence, from Eq. \eqref{eq:51}, this means $g_{1}=g_{2}=g$; using this values in Eqs. \eqref{eq:50} to \eqref{eq:54} we deduce
\begin{equation}
\mathcal{E}_{2}^{DS}=\mathcal{E}_{2}^{\left(A_{1}|A_{3}\right)}=\mathcal{E}_{2}^{\left(A_{2}|A_{3}\right)}. \label{eq:56}
\end{equation}
Where the superscript $DS$ in Eq. \eqref{eq:56} means `Detector-System'; it is used to denote  the R{\'e}nyi-2 entanglement of the reduced bipartition including the mode of any detector and the Gaussian system under measurement.  We carry out a numerical minimization for the differences $\sqrt{a^{2} + a_{3}^2 - 1}- a$ and $\sqrt{a^{2} + a_{3}^2 - 1}- \alpha$, $ ~\forall~r,\theta$ obtaining respectively $\approx 1.464$ and $1.428$; this implies that $\sqrt{a^{2} + a_{3}^2 - 1} > \left\lbrace a, \alpha \right\rbrace$; therefore, according to Eqs. \eqref{eq:50} and \eqref{eq:51}, we have $\mathcal{E}_{2}^{DS} \neq 0$. On the other hand, we analyze the difference $a-\alpha$ as a function of $r$ and $\theta$, finding the two cases: $\alpha < a$ and $a \leq \alpha$; therefore, according to Eqs. \eqref{eq:50} and \eqref{eq:51} we have 
\begin{equation}
\mathcal{E}_{2}^{DS}=\frac{1}{2}\ln g, \label{eq:57}
\end{equation}
with
\begin{equation}
g=\begin{cases}
    \frac{\beta}{8 a^2}, & \text{if $\alpha< a$},\\
    \left(\frac{a^2 - a_{3}^2}{a^2 - 1}\right)^2, & \text{if $a \leq  \alpha.$}
  \end{cases} \label{eq:58}
\end{equation}
\begin{figure}
\includegraphics[width=0.47\textwidth]{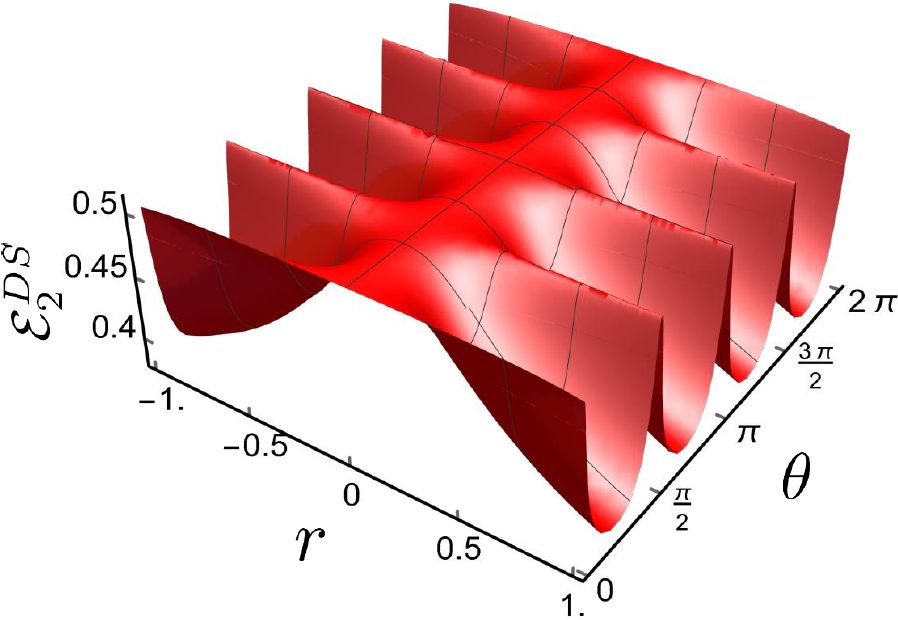}
\centering
\caption{Plot for the reduced entanglement $\mathcal{E}_{2}^{DS}$ as a function of $r$ and $\theta$. This quantity represent the amount of entanglement contained in the reduced bipartition containing the mode of any detector and that of the Gaussian system under inspection. It has an oscillatory behaviour with period $\pi/2$. The maximum value is $\ln \left(5/3 \right)$ attained when the Gaussian system under measurement is a minimum uncertainty state. The minimum values are at $\theta=(2n + 1)\pi/4,~n= 0, 1, 2,  \cdots $ for $r\neq 0$, tending asymptotically to the absolute minimum of $(1/2) \ln(2)$ in the limit situation of $\left|r\right| \longrightarrow \infty$.}\label{fig:5}
\end{figure}

What essentially Eq. \eqref{eq:55} means is that the partie $A_{3}$, which contains the mode of the Gaussian system under measurement, is the principal support of the tripartite entanglement relationship between $A_{1}, A_{2}$ and $A_{3}$; that is, without the partie $A_{3}$, there is no possibility to see any entanglement in the reduced system containing the parties $A_{1}$ and $A_{2}$ only, which is an expected fact since, under the interaction Hamiltonian, Eq. \eqref{eq:1}, one of the two conjugate observables of the detectors is directly linked with one of the two belonging to the Gaussian system under inspection, just as we sketch in Fig. \ref{fig:1}. Therefore,  the entanglement relation between the detectors in the tripartite system is an inherent consequence of their coupling with the Gaussian system under observation. On the other hand, the Eq. \eqref{eq:56} imply that the two reduced bipartitions containing the modes of any detector and the Gaussian system under measurement present the same amount of entanglement for the same $r$ and $\theta$. 
Notably, these reduced entanglements are distinct from zero, which is a necessary fact for the information transfer of the canonical pair on study toward the quantum state of the detectors; this behaviour is again a consequence of the symmetry of the  Hamiltonian, Eq. \eqref{eq:1}. In Fig. \ref{fig:5} we plot $\mathcal{E}_{2}^{DS}$ as a function of $r$ and $\theta$. 
\begin{figure}
\includegraphics[width=0.47\textwidth]{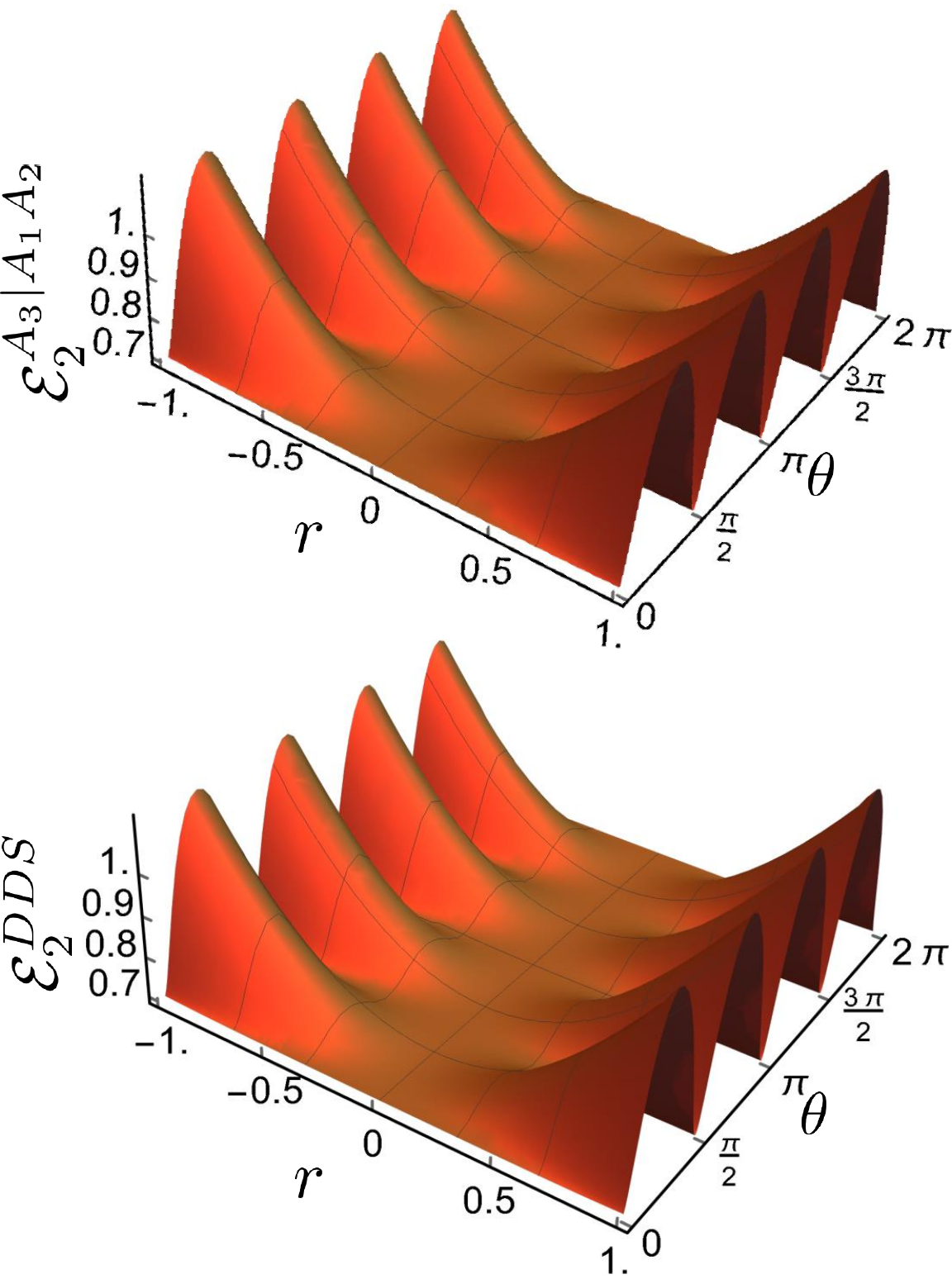}
\centering
\caption{Plots for the R{\'e}nyi-2 entanglements $\mathcal{E}_{2}^{DDS}$ (bottom figure) and $\mathcal{E}_{2}^{A_{3}|A_{1}A_{2}}$ (upper figure) as functions of $r$ and $\theta$. Such quantities dictate the amount of entanglement for the global bipartitions focused on the partie containing the mode of any detector and that of the Gaussian system under measurement respectively. They have an oscillatory behaviour with a period of $\pi/2$. Their maximum values are reached at $\theta= \left(2n + 1\right)\pi/4,~~n=0,1,2, \cdots$ for any $r\neq 0$. Besides, they are increasing functions of $\left|r \right|$. The minimum values are $\text{min}\left[\mathcal{E}_{2}^{DDS} \right]=(1/2)\ln(4)$ and $\text{min}\left[\mathcal{E}_{2}^{A_{3}|A_{1}A_{2}} \right]=(1/2)\ln(9)$, reached when the Gaussian system under measurement is a minimum uncertainty state.}\label{fig:8}
\end{figure}

\begin{figure}
\includegraphics[width=0.47\textwidth]{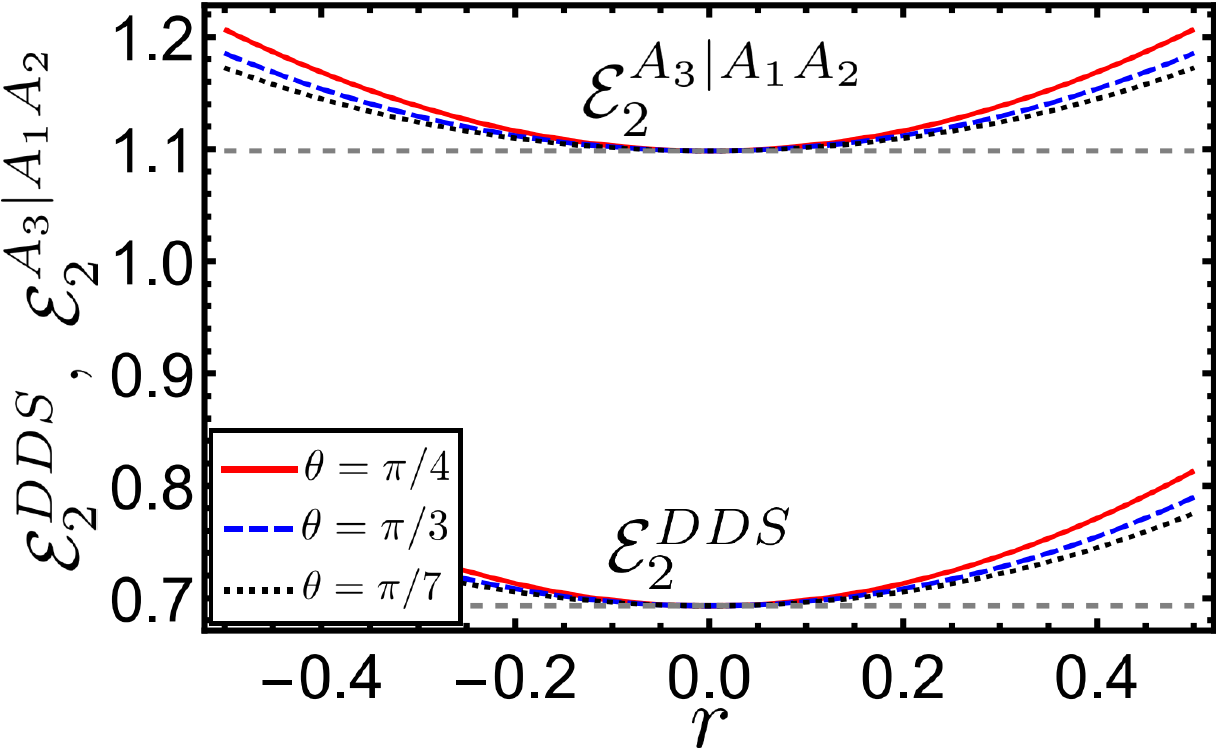}
\centering
\caption{Plots for the R{\'e}nyi-2 entanglements $\mathcal{E}_{2}^{DDS}$ and $\mathcal{E}_{2}^{A_{3}|A_{1}A_{2}}$ (bottom and upper curves respectively) at various fixed rotation angles. The gray dashed lines represent the possible minimum values for these quantities. From this, we can see that $\mathcal{E}_{2}^{A_{3}|A_{1}A_{2}}>\mathcal{E}_{2}^{DDS}.$}\label{fig:8.1}
\end{figure}

Taking into account Eqs. \eqref{eq:45} and \eqref{eq:54.1} and the considerations for the reduced two-mode R{\'e}nyi-2 entanglements explained before, we have the following expressions for all $\mathcal{E}_{2}^{(A_{i}|A_{j} A_{k})}$ of the system
\begin{equation}
\mathcal{E}_{2}^{DDS}=\mathcal{E}_{2}^{(A_{1,2}|A_{2,1} A_{3})} =\frac{1}{2} \ln (a^2),\label{eq:58.1}
\end{equation}
\begin{equation}
\mathcal{E}_{2}^{(A_{3}|A_{1} A_{2})} =\frac{1}{2} \ln (a_{3}^2), \label{eq:58.2}
\end{equation}
where the superscript \textit{DDS} in Eq. \eqref{eq:58.1} stand as `Detector, Detector-System'; it is used to denote the R{\'e}nyi-2 entanglements of the global bipartitions containing any mode detector as focus. Hence, Eq. \eqref {eq:58.1}, imply that any of the two global bipartitions focused on the partie containing the mode of any detector, present the same amount of entanglement for the same $r$ and $\theta$ parameters; this entanglement is ever distinct from zero since the minimum value of $a_{1}$ and $a_{2}$ (consequently of $a$) is $2$ as can be directly verified from Eqs. \eqref{eq:46} and \eqref{eq:47}. This symmetry is again a consequence of the linear interaction, Eq. \eqref{eq:1}, linking symmetrically the system of the two detectors with the Gaussian system under observation. In Fig. \ref{fig:8} we plot $\mathcal{E}_{2}^{DDS}$ and $\mathcal{E}_{2}^{\left(A_{3}| A_{1} A_{2}\right)}$ as functions of $r$ and $\theta$. From Eqs. \eqref{eq:46} to \eqref{eq:48} we can verify $\left\lbrace a_{1}, a_{2}, a_{3}\right\rbrace > 1$ and $a_{3}> \left\lbrace a_{1}, a_{2}\right\rbrace$ for a determined $r$ and $\theta$; then, from Eqs. \eqref{eq:58.1} and \eqref{eq:58.2} we deduce $\mathcal{E}_{2}^{\left(A_{3}|A_{1}A_{2} \right)}> \mathcal{E}_{2}^{DDS}$; see Fig \ref{fig:8.1}.

\begin{figure}
\includegraphics[width=0.47\textwidth]{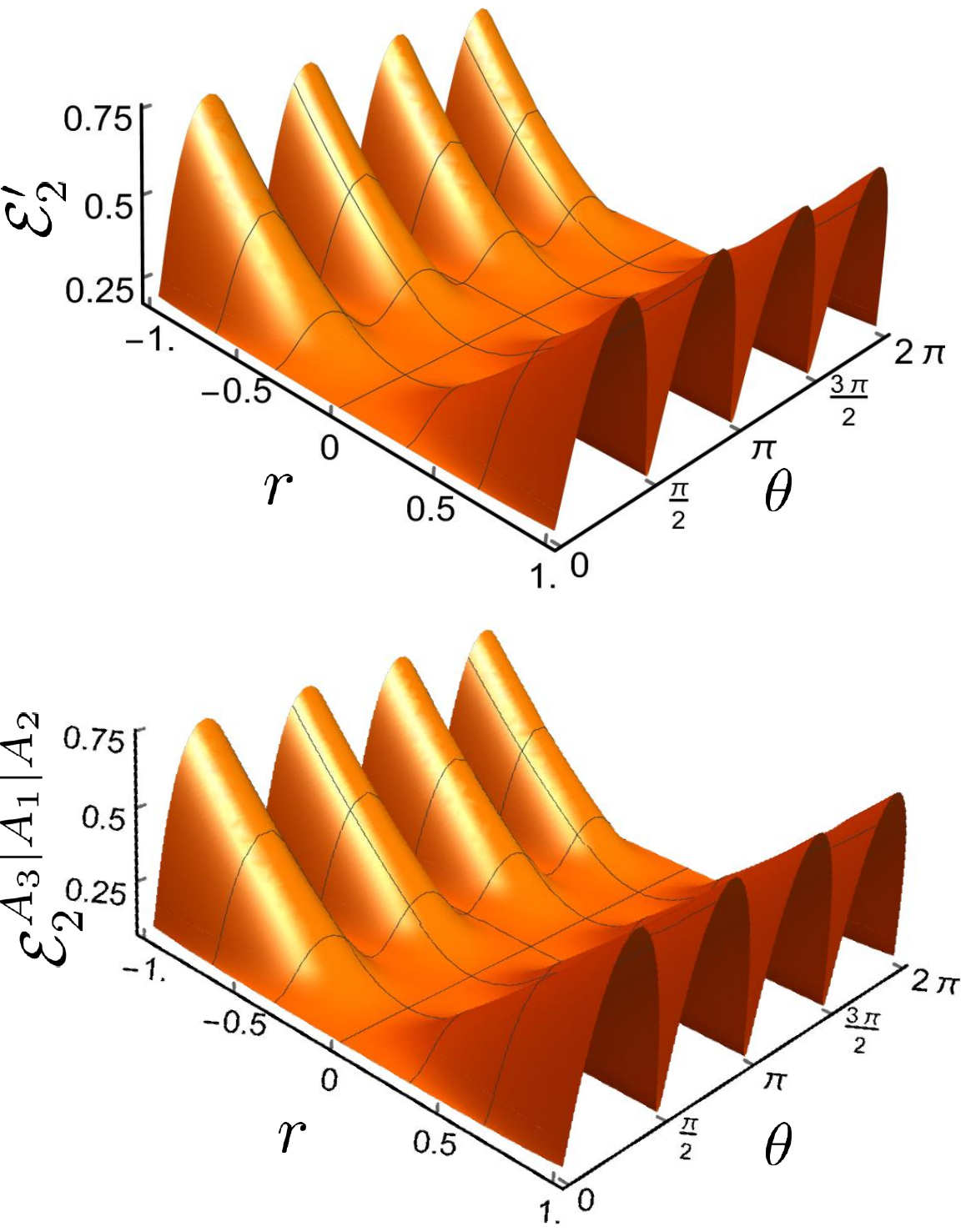}
\centering
\caption{Plot for the residual tripartite entanglements $\mathcal{E}_{2}'$ (top figure) and $\mathcal{E}_{2}^{\left(A_{3}| A_{1}| A_{2}\right)}$ (bottom figure) as functions of $r$ and $\theta$. These quantities quantify the complete amount of tripartite entanglement in our Gaussian Arthurs-Kelly measurement setting. They are focused respectively on any detector and on the Gaussian system under examination. They have an oscillatory behavior with period $\pi/2$. They have a maximum value for a determined $r\neq 0$ at $\theta= (2n + 1)\pi/4,~n=0,1,2, \cdots$. Besides, the global tripartite entanglement grow as the squeezing factor $r$ also grows. The minimum values are respectively $\ln(6/5)$ and $\ln (27/25)$, attained when the system under measurement is a minimum uncertainty state.}\label{fig:6}
\end{figure}
\begin{figure}
\includegraphics[width=0.47\textwidth]{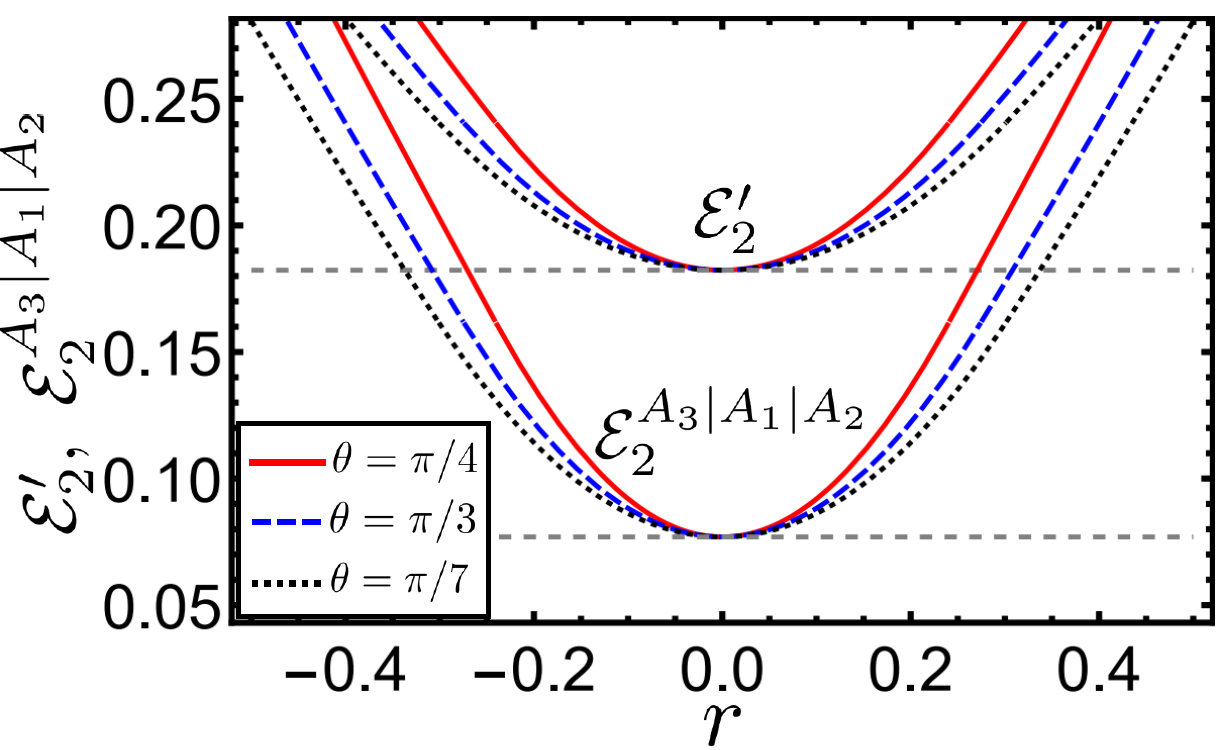}
\centering
\caption{Plots for the tripartite entanglements $\mathcal{E}_{2}'$ and $\mathcal{E}_{2}^{A_{3}|A_{1}|A_{2}}$  (top and lower curves respectively) at various fixed rotation angles. The gray dashed lines represent the possible minimum values for these quantities. From this, we can see that $\mathcal{E}_{2}'>\mathcal{E}_{2}^{A_{3}|A_{1}|A_{2}}.$}\label{fig:7}
\end{figure}

For the full tripartite entanglements, we need to consider the three focus options for the expression, Eq. \eqref{eq:49}; however, by taking into account the same considerations that for the reduced two-mode R{\'e}nyi-2 entanglements, it is straightforward verify that
\begin{equation}
\mathcal{E}_{2}'=\mathcal{E}_{2}^{\left(A_{1}| A_{2}| A_{3}\right)}=\mathcal{E}_{2}^{\left(A_{2}| A_{1}| A_{3}\right)}; \label{eq:59}
\end{equation}
then, using Eqs. \eqref{eq:50} and \eqref{eq:51} we have
\begin{equation}
\mathcal{E}_{2}'=\frac{1}{2}\ln\left(\frac{a^2}{g} \right). \label{eq:60}
\end{equation}
In a similar procedure we get
\begin{equation}
\mathcal{E}_{2}^{\left(A_{3}| A_{1}| A_{2}\right)}=\frac{1}{2}\ln \left( \frac{a_{3}^2}{g^2}\right). \label{eq:61}
\end{equation}
According to Eqs. \eqref{eq:60} and \eqref{eq:61}, the amount of tripartite entanglement is the same when the focus is the partie containing the mode of any detector and distinct from the case when is focused on the mode of the Gaussian system under examination. In Fig. \ref{fig:6} we plot $\mathcal{E}_{2}'$ and $\mathcal{E}_{2}^{\left(A_{3}| A_{1}| A_{2}\right)}$ as functions of $r$ and $\theta$.  Both tripartite entanglements  grow as the magnitude of the squeezing of the Gaussian system under observation increases, in concordance with a loss of purity of the reduced single-mode systems, i.e., a loss of information about the knowledge of its initial quantum states.  Notably, the tripartite entanglement focused on any detector is greater than the one focused on the Gaussian system under inspection for fixed $r$ and $\theta$ values, then $\mathcal{E}_{2}' >\mathcal{E}_{2}^{\left(A_{3}|A_{1}|A_{2}\right)}$; see Fig. \ref{fig:7}. This behavior is consistent with the description of a user locally localized in the partie (focus) containing a single mode with associated symplectic invariant $a_{i}$. Then, the entanglement from the perspective of this observer will be like its total `at sight' of the single-mode invariants associated with the other two parties $a_{j}$ and $a_{k}$; then due to Eqs.  \eqref{eq:46} to Eq. \eqref{eq:48},  the observer will see a higher amount of tripartite entanglement from the perspective of the partie containing the mode of any detector.


\section{Conclusions} \label{Sec. 4}
This contribution uncovers key features of the simultaneous measurement process --of the position and momentum observables-- proposed by Arthurs and Kelly. Particularly, when the system under inspection is a rotated, displaced, and squeezed vacuum state,  we have a measurement configuration entirely made up of Gaussian states and the dynamics of the measurement process generates tripartite continuous variable entanglement.

Through the PPT criterion, we determine the non-validity of each of the three partially transposed covariance matrices associated with the three $i|jk$-mode bipartition of the system. In particular, we find that a couple of symplectic eigenvalues of each transposed covariance matrix is less than one, which is equivalent to the violation of the Heisenberg uncertainty relation for multiple modes: $\boldsymbol \sigma + i \boldsymbol \Omega \geq 0$. Then, we conclude that the system is entangled between the three modes, i. e., the two detectors and the Gaussian system under observation throughout the dynamics of the measurement process, being this aspect independent of the squeezing, rotation, and displacement of the Gaussian system under observation. This fact allows categorizing the generated Gaussian tripartite entanglement in the class of fully inseparable states according to the classification exposed in \cite{Giedke2001}. 

Besides, we study the quantitative entanglement properties of the system by using the residual tripartite R{\'e}nyi-2 entanglement as a quantifier measure. We find that the angle of rotation $\theta$ and the squeezing $r$ of the Gaussian system under measurement affect the amount of entanglement in all reduced and global mode bipartition, as well as the complete tripartite entanglement of the system; therefore, the initial preparation of such system gets involved in the quantitative entanglement properties developed throughout the dynamics of the measurement process. 

We find a symmetric entanglement structure in the mode bipartitions of the system. In particular, the reduced bipartition containing only the modes of the two detectors has zero entanglement; this means that the party associated with the Gaussian system under measurement is the principal support of the tripartite entanglement relation in the measurement setting. That is, as a consequence of the symmetry associated with the interaction Hamiltonian, Eq. \eqref{eq:1}, the canonical pair of the Gaussian system under measurement gets linked with the canonical set composed by an observable of the first detector and the corresponding conjugate variable of the second detector; see Fig. \ref{fig:1}. Also, as a consequence of the symmetry of the Eq. \eqref{eq:1}, the two reduced bipartitions that contain the modes of the Gaussian system under measurement and the one of a detector, present the same quantity of entanglement, which implies that both detectors are equally entangled with the Gaussian system under observation; then, this is the fact that allows the transfer of information about the canonical pair under measurement to the quantum state of the detectors.

On the other hand, the two global bipartitions focused on any detector contain an equal quantity of entanglement, which comes again from the symmetry of the interaction Hamiltonian given by Eq. \eqref{eq:1}. Notably, the global bipartition focused on the mode of the Gaussian system under observation presents a higher quantity of entanglement from those focused on the detectors. Therefore, the amount of entanglement in any global bipartition of the system will depend
logarithmically on the symplectic invariant $a_{i}$ associated with the mode chosen as focus, as Eqs. \eqref{eq:58.1} and \eqref{eq:58.2} suggests. 

The full tripartite entanglements quantified by the  residual tripartite R{\'e}nyi-2 entanglement depend, in general, on the party chosen as the focus. We find the same amount of entanglement when the focus is the party containing the mode of any detector, which is again a consequence of the symmetry of the interaction Hamiltonian governing the dynamics of the measurement process. Besides, the tripartite entanglement focused on any detector is greater than the one focused on the Gaussian system under examination. This behavior is consistent with the description of an observer localized on the party chosen as focus, where the entanglement from the perspective of this observer will be like `at sight' of the reciprocal purities associated with the other two parties.

\begin{acknowledgments}
J. A. Mendoza-Fierro thanks CONAHCYT for the
postdoctoral fellowship support under the application number 3762623.
\end{acknowledgments}

\appendix

\section{ \label{appendixA}Block components of the covariances matrices}
In this appendix, we define the $2 \times 2$ dimensional block matrices composing the CMs, Eqs. \eqref{eq:34a} to \eqref{eq:34c}. All terms appearing depend on the parameters $r$ and $\theta$; in the subsequent, we omit this dependence for brevity. Then,  we have ($\alpha_{1}=\alpha_{2}=1$)
\begin{equation}
\boldsymbol \sigma_{1} = \begin{bmatrix}
\begin{array}{cc}
\delta_{\hat{q}_{1}}^2 + \left(\delta_{\hat{p}_{2}}/2\right)^2 + \delta_{\hat{q}_{3}}^2& 0 \\
 0 &\delta_{\hat{p}_{1}}^2 \\
\end{array}
  \end{bmatrix},
\end{equation}
\begin{equation}
\boldsymbol \sigma_{2} = \begin{bmatrix}
\begin{array}{cc}
\left(\delta_{\hat{p}_{1}}/2\right)^2 + \delta_{\hat{q}_{2}}^2 + \delta_{\hat{p}_{3}}^2& 0 \\
 0 &\delta_{\hat{p}_{2}}^2 \\
\end{array}
  \end{bmatrix},
\end{equation}
\begin{equation}
\boldsymbol \sigma_{3} = \begin{bmatrix}
\begin{array}{cc}
  \delta_{\hat{p}_{2}}^2 + \delta_{\hat{q}_{3}}^2  & \text{Cov}\\
  \text{Cov} & \delta_{\hat{p}_{1}}^2 + \delta_{\hat{p}_{3}}^2\\
\end{array}
  \end{bmatrix},
\end{equation}
\begin{equation}
\boldsymbol \varepsilon_{1, 2} = \begin{bmatrix}
    \begin{array}{cc}
 \text{Cov} & \frac{\delta_{\hat{p}_{2}}^2}{2} \\
 -\frac{\delta_{\hat{p}_{1}}^2}{2} & 0 \\
\end{array}
  \end{bmatrix},
\end{equation}
\begin{equation}
\boldsymbol \varepsilon_{1, 3} = \begin{bmatrix}
   \begin{array}{cc}
\frac{\delta_{\hat{p}_{2}}^2}{2} + \delta_{\hat{q}_{3}}^2  & \text{Cov} \\
 0 & -\delta_{\hat{p}_{1}}^2 \\
\end{array}
  \end{bmatrix},
\end{equation}
\begin{equation}
\boldsymbol \varepsilon_{2, 3} = \begin{bmatrix}
  \begin{array}{cc}
 \text{Cov} &\frac{\delta_{\hat{p}_{1}}^2}{2} + \delta_{\hat{p}_{3}}^{2}\\
\delta_{\hat{p}_{2}}^2 & 0 \\
\end{array}
  \end{bmatrix};
\end{equation}
where the variances $\delta_{\hat{X}}^2$ with $\hat{X}\in \left\lbrace \hat{x}_{1}, \hat{x}_{2}, \hat{x}_{3}, \hat{p}_{1}, \hat{p}_{2}, \hat{p}_{3} \right\rbrace$, were defined in Sec. \ref{subsec:l}. On the other hand, the term $\text{Cov}$ represent the covariance of the Gaussian system under observation, this is 
\begin{equation}
\text{Cov}=\sin (2 \theta ) \left(e^{2r}- e^{-2r} \right).
\end{equation}
Besides, we recall that the balance parameter appearing in $\delta_{\hat{x}_{j}}^2$ and $\delta_{\hat{p}_{j}}^2$ with $j=1,2$, is fixed at the rate $b=\delta_{\hat{x}_{3}}/\delta_{\hat{p}_{3}}$ as we explain in Sec. \ref{subsec:l}.
\bibliographystyle{apsrev4-2}
\bibliography{bibliography}
\end{document}